\documentclass[preprint,3p,12pt]{elsarticle}
\usepackage{mathrsfs}
\usepackage{amsmath}
\usepackage{stmaryrd}
\usepackage{bbding}
\usepackage{dcolumn}
\usepackage{graphicx}
\usepackage{amsfonts}
\usepackage{amssymb}
\usepackage{psfrag}
\usepackage{wrapfig}
\usepackage{subfigure}
\usepackage{makeidx}
\usepackage{bm}
\usepackage{epsf}
\usepackage{epsfig}
\usepackage{setspace}
\usepackage{graphicx}
\usepackage{epstopdf}
\usepackage{psfrag}
\usepackage{subfigure}
\usepackage{color}

\begin{document}

\title{A Well-Balanced Unified Gas-Kinetic Scheme for Multiscale Flow Transport Under Gravitational Field}

\author[PKU]{Tianbai Xiao}
\ead{xiaotianbai@pku.edu.cn}

\author[PKU]{Qingdong Cai}
\ead{caiqd@pku.edu.cn}

\author[HKUST]{Kun Xu\corref{cor1}}
\ead{makxu@ust.hk}

\address[PKU]{Department of Mechanics and Engineering Science, College of Engineering, Peking University, Beijing 100871, China}
\address[HKUST]{Department of Mathematics, Department of Mechanical and Aerospace Engineering, Hong Kong University of Science and Technology, Clear Water Bay, Kowloon, Hong Kong}
\cortext[cor1]{Corresponding author}

\begin{abstract}

The gas dynamics under gravitational field is usually associated with the multiple scale nature due to large density variation and a wide range of
local Knudsen number.
It is challenging to construct a reliable numerical algorithm to accurately capture the non-equilibrium physical effect in different regimes.
In this paper, a well-balanced unified gas-kinetic scheme (UGKS) for all flow regimes under gravitational field will be developed, which can be used for the study of non-equilibrium gravitational gas
system.
The well-balanced scheme here is defined as a method  to evolve an isolated gravitational system under any initial condition to
an isothermal hydrostatic equilibrium state and to keep such a solution.
To preserve such a property  is important for a numerical scheme,  which can be used for the study of slowly evolving gravitational system, such as the formation of star and galaxy.
Based on the Boltzmann model with external forcing term, an analytic time evolving (or scale-dependent) solution is constructed to provide the corresponding dynamics in the
cell size and time step scale, which is subsequently used in the construction of UGKS. As a result, with the variation of the ratio between the numerical time step and local particle collision time,
the UGKS is able to recover flow physics in different regimes and provides a continuum spectrum of gas dynamics.
For the first time, the flow physics of a gravitational system in the transition regime  can be studied using the UGKS, and
the non-equilibrium phenomena in such a gravitational system can be clearly identified.
Many numerical examples will be used to validate the scheme.
New physical observation, such as the correlation between the gravitational field and the heat flux in the transition regime, will
be presented. The current method provides an indispensable tool for the study of non-equilibrium gravitational system.

\end{abstract}


\maketitle

\section{Introduction}

The universe is an evolving gravitational system.
The gas dynamics due to the gravitational force plays a critical role in the star and galaxy formation, as well as atmospheric convection on the planets.
For an evolving gravitational system, there is almost no any validated governing equation to describe the non-equilibrium dynamics uniformly across different regimes.
The well-established statistical mechanics is mostly for the equilibrium solution.
For an isolated gravitational system, under the conservation of mass, momentum, and energy,
the system will eventually get to the isothermal equilibrium state from an arbitrary initial condition.
Such a steady-state solution will be maintained due to
exact balance between the gravitational source term and the inhomogeneous flux function.
For the study of a gravitational system, many numerical schemes have been developed.
To capture such an equilibrium solution for an isolated gravitational system is a minimal requirement for a scheme which can be used in the astrophysical applications.
The lack of well-balanced property may lead to spurious solution, or even present wrong physical solution for a long time evolving gravitational system.
The purpose of this paper is to develop such a scheme which not only has the well-balanced property, but also be able to capture the non-equilibrium phenomena,
which have not been studied theoretically or numerically for a gravitational system before.

For the equilibrium flow, such as for the gravitational Euler system, many efforts have been devoted to the construction of well-balanced schemes.
Leveque and Bale \cite{leveque1999wave} developed a quasi-steady wave-propagation algorithm, which is able to capture perturbed quasi-steady solutions.
Botta et al. \cite{botta2004well} used local, time dependent hydrostatic reconstructions to achieve the hydrostatic
balance. Xing and Shu \cite{xing2013high} proposed a high-order WENO scheme to resolve small perturbations on the hydrostatic balance state on coarse meshes.
The basic idea of these methods is to extend classical algorithm for the compressible Euler equations to the gravitational system with proper modifications.

Different gas dynamic equations can be constructed under different modeling scales.
In the kinetic scale, such as particle mean free path and traveling time between particle collisions, the Boltzmann
equation is well established.
In the hydrodynamic scale, even though the scale for its modeling is not clearly identified, the Euler and Navier-Stokes
equations are routinely used. Due to the clear scale separation, both kinetic and hydrodynamic equations can be applied in their respective scales.
However, the real gas dynamics may not have such a scale separation. With the scale variation between the kinetic and hydrodynamic ones,
the gas dynamic equation should have a smooth transition from the Boltzmann to the Navier-Stokes equations.
This multiple scale nature is especially important for a gravitational system, where due to the gravitational effect
the density in the system can be varied largely, and so is the particle mean free path. With a fixed modeling scale,
such as the mesh size in a numerical scheme,
the cell's Knudsen number $Kn_c$ can be changed significantly.
For a single gravitational system, the flow dynamics may vary from the
kinetic Boltzmann modeling in the upper atmospheric layer to the hydrodynamic one in the inner high density region, with a continuous variation of flow physics.
Therefore, a gravitational system has an intrinsic multiple scale nature.
The corresponding numerical algorithm to simulate such a system is preferable to have a property of  providing a continuum spectrum of flow dynamics from rarefied to continuum one.

In recent years, the unified gas-kinetic scheme has been developed for the simulation of multiple scale flow problems \cite{xu2010unified,huang2012unified,huang2013unified,xubook}.
This algorithm is based on the direct physical modeling on the mesh size scale, such as constructing the corresponding governing equations in such a scale.
The scheme is able to capture physical solution in  all flow regimes.
The coupled treatment of particle transport and collision in the evaluation of a time-dependent interface flux function is the key for its cross-scale  modeling
and ensures a multiple scale nature of the algorithm.
The mechanism of flow evolution in different regimes is determined by the ratio between the time step and the local particle collision time.
In this paper, the further development of the scheme for a gravitational system is proposed.

In order to develop a well-balanced gas-kinetic scheme, the most important ingredient is to take the external force effect into the flux transport across a cell interface.
Attempts have been made in the construction of the schemes for the shallow water equations \cite{xushallow} and gas dynamic equations \cite{tian2007three,xuluochen,luo2011well}.
In this paper, the similar methodology is used in the unified gas kinetic scheme.
The scheme can be used to study the multiple scale non-equilibrium flow phenomena under gravitational field, which has never been fully explored before.

This paper is organized as following.
Section 2 is about the kinetic theory under gravitational field.
Section 3 presents the construction of the unified gas-kinetic scheme for a gravitational system.
Section 4 includes numerical examples to demonstrate the performance of the scheme. The last section is the conclusion.

\section{Gas kinetic modeling}

The gas kinetic theory describes the evolution of particle distribution function $f(x_i,t,u_i,\xi)$ in space and time
$(x_i,t)$. Here $u_i=(u,v,w)$ is particle velocity and $\xi$ is the internal variable for the rotation and vibration.
The evolution equation for $f$ in the kinetic scale with the separate modeling of particle transport and collision  is the so-called Boltzmann Equation,
\begin{equation*}
	f_t+u_i f_{x_i}+\phi_if_{u_i}=Q(f,f),
\end{equation*}
where $\phi_i$ is the external forcing term and $Q(f,f)$ is the collision term.

Many kinetic models, such as BGK \cite{bhatnagar1954model}, ES-BGK \cite{holway1966new}, Shakhov \cite{shakhov1968generalization},
and even the full Boltzmann equation \cite{liu2016unified}, can be used in the construction of the unified gas kinetic scheme.
Here we use the model equation to explain the principle for the algorithm development.
The generalized one-dimensional BGK-type model with the inclusion of external force $\phi_x$ can be written as
\begin{equation}
f_t+uf_x+\phi_x f_u=\frac{f^+-f}{\tau},
\label{eqn:1D external force}
\end{equation}
where $f^+$ is the equilibrium state, and $\tau$ is the collision time.
For the BGK equation, $f^+$ is exactly the Maxwellian distribution
\begin{equation*}
f^+=g_0=\rho\left(\frac{\lambda}{\pi}\right)^\frac{K+1}{2}e^{-\lambda[(u-U)^2+\xi^2]},
\end{equation*}
where $\lambda={\rho}/{(2p)}$ and K is the dimension of $\xi$.
For the Shakhov model equation, $f^+$ takes the form,
\begin{equation*}
f^+=g_0\left[1+(1-\mathrm{Pr} )c q\left(\frac{c^2}{RT}-5\right)/(5pRT)\right],
\end{equation*}
where $c=u-U$ is the peculiar velocity, $q$ is heat flux,  and Pr is Prandtl number.
The collision term satisfies the compatibility condition
\begin{equation*}
\int (f^+-f)\psi d\Xi=0,
\end{equation*}
where $\psi=\left(1,u,\frac{1}{2}(u^2+\xi^2) \right)^T$ is a vector of moments for collision invariants,
and $d\Xi=dud\xi$. The macroscopic conservative flow variables are the moments of the particle distribution function via
\begin{equation*}
\textbf{W} =\left(
\begin{matrix}
\rho &\\
\rho U &\\
\rho E &
\end{matrix}
\right)=\int f\psi d\Xi.
\end{equation*}

With a local constant collision time $\tau$, the integral solution of Eq.(\ref{eqn:1D external force}) can be constructed by the method of characteristics,
\begin{equation}
\begin{aligned}
f(x,t,u,\xi)=&\frac{1}{\tau}\int_{t^n}^t f^+(x',t',u',\xi)e^{-(t-t')/\tau}dt' \\
&+e^{-(t-t^n)/\tau}f_0^n(x^n,t^n,u^n,\xi),
\end{aligned}
\label{eqn:integral solution}
\end{equation}
where $x'=x-u'(t-t')-\frac{1}{2}\phi_x (t-t')^2$ and $u'=u-\phi_x (t-t')$ are the trajectories in physical and phase space,
and $f_0^n$ is the gas distribution function at the beginning of $n$-th time step.
The above integral solution plays the most important role for the construction of the well-balanced UGKS.

\section{Numerical algorithm}
\subsection{Construction of interface distribution function}

In the unified scheme,  the distribution function $f({x_{i+1/2},u_k})$ at the cell interface $x_{i+1/2}$
is constructed from the evolution solution Eq.(\ref{eqn:integral solution}) for the
interface flux evaluation.
With the notation of $x_{i+1/2}=0$ and $t^n=0$, the time-dependent interface distribution function becomes
\begin{equation}
\begin{aligned}
	f(0,t,u_k,\xi)=&\frac{1}{\tau}\int_0^t f^+(x',t',u_k',\xi)e^{-(t-t')/\tau}dt' \\
	&+e^{-t/\tau}f_0(x^0,0,u_{k}^0,\xi)
\end{aligned}
\end{equation}
where $(x^0,u^0)$ is the initial location in physical and velocity space for the particle which passes through the cell interface at time $t$.
Based on the above integral solution with particle acceleration, the well-balanced algorithm can be constructed similarly  as the original UGKS.
Note that time accumulating effect from the external forcing term on the time evolution of the gas distribution function should be explicitly taken into account.

To the second-order accuracy, the initial gas distribution function $f_0$ around the cell interface $x_{i+1/2}$ is reconstructed as
\begin{equation*}
f_0(x,0,u_k,\xi)=\left\{
\begin{aligned}
&f_{i+1/2,k}^L+\sigma_{i,k}x, \quad x\le 0 ,\\
&f_{i+1/2,k}^R+\sigma_{i+1,k}x, \quad x> 0 ,
\end{aligned}
\right.
\end{equation*}
where $f_{i+1/2,k}^L$ and $f_{i+1/2,k}^R$ are the reconstructed initial distribution functions at the left and right hand sides of the cell interface.
In the current scheme, the van Leer limiter is used in the reconstruction.

The equilibrium distribution function around a cell interface is approximated locally through a Taylor expansion in space and time as
\begin{equation}
g=g_0\left[1+(1-H[x]){a}^Lx+H[x]{a}^Rx+{A}t\right],
\label{eqn:equilibrium expansion}
\end{equation}
where $g_0$ is the Maxwellian distribution at $(x=0, t=0)$, and $H[x]$ is the Heaviside step function.
Here $a^L,a^R$, and $A$, are from the Taylor expansion of a Maxwellian,
\begin{equation*}
\begin{aligned}
&a^{L,R}=a_1^{L,R}+a_2^{L,R}u+a_3^{L,R}\frac{1}{2}(u^2+\xi ^2)=a_{\alpha}^{L,R}\psi_\alpha, \\
&A=A_1+A_2u+A_3\frac{1}{2}(u^2+\xi ^2)=A_{\alpha}\psi_\alpha.
\end{aligned}
\end{equation*}

Based on the compatibility condition at the cell interface, the equilibrium and the corresponding macroscopic conservative variables $\textbf{W}_0$ can be determined from
\begin{equation*}
\int (f^+-f)|_{x=0,t=0}\psi d\Xi=0,
\end{equation*}
where $d\Xi=dud\xi$, which results
\begin{equation*}
\int g_0\psi_\alpha d\Xi=\textbf{W}_0=\sum_{u_k>0}f_{i+1/2,k}^L\psi d\Xi+\sum_{u_k<0} f_{i+1/2,k}^R\psi d\Xi.
\end{equation*}

After the determination of the equilibrium state at the cell interface, its spatial slopes $a^L,a^R$ can be obtained from the slopes of conservative variables on both sides of a cell interface.
\begin{equation*}
\left(\frac{\partial {\textbf{W}}}{\partial x}\right)^L=\int a^Lg_0\psi d\Xi, \quad \left(\frac{\partial {\textbf{W}}}{\partial x}\right)^R=\int a^Rg_0\psi d\Xi.
\end{equation*}

The time derivative $A$ of $g_0$ is related to the temporal variation of conservative flow variables
\begin{equation*}
\frac{\partial {\textbf{W}}}{\partial t}=\int Ag_0\psi d\Xi,
\end{equation*}
 and it can be calculated via time derivative of the compatibility condition
\begin{equation*}
\frac{d}{dt}\int (f^+-f)\psi d\Xi \mid_{x=0,t=0}=0.
\end{equation*}
With the help of the Euler equations with external forcing term, it gives
\begin{equation*}
-\int u\frac{\partial g}{\partial x}\psi d\Xi-\int \phi_x \frac{\partial g}{\partial u}\psi d\Xi=\frac{\partial\textbf{W}}{\partial t}=\int Ag_0\psi d\Xi ,
\end{equation*}
where the spatial derivative can be constructed from the Taylor expansion of equilibrium distribution Eq.(\ref{eqn:equilibrium expansion}),
 and the velocity derivative is obtained from the exact Maxwellian distribution. The result is
\begin{equation*}
-\int a^{L,R}ug_0\psi d\Xi+2\int \phi_x \lambda_0 g_0 (u-U_0)\psi d\Xi =\frac{\partial\textbf{W}}{\partial t}=\int Ag_0\psi d\Xi,
\end{equation*}
where $U_0$ and $\lambda_0$ are the corresponding macroscopic variables in the equilibrium state $g_0$. Using the equation above, we can obtain coefficients $A=(A_1,A_2,A_3)^T$.

After the determination of all coefficients,  the time dependent interface distribution function becomes
\begin{equation*}
\begin{aligned}
f(0,t,u_k,\xi)=&\left(1-e^{-t/\tau}\right)(g_0'+g'^+)\\
&+\left(\tau(-1+e^{-t/\tau})+te^{-t/\tau}\right)a^{L,R}u_kg_0' \\
&-\left[\tau\left(\tau(-1+e^{-t/\tau})+te^{-t/\tau}\right)+\frac{1}{2}t^2e^{-t/\tau}\right] a^{L,R}\phi_x g_0' \\
&+\tau \left(t/\tau-1+e^{-t/\tau}\right){A}g_0' \\
&+e^{-t/\tau}\left[\left(f_{i+1/2,k^0}^L+\left(-(u_k-\phi_xt)t-\frac{1}{2}\phi_xt^2\right)\sigma_{i,k^0}\right)H[u_k-\frac{1}{2}\phi_x t] \right.\\
&\left. +\left(f_{i+1/2,k^0}^R+\left(-(u_k-\phi_x t)t-\frac{1}{2}\phi_x t^2\right)\sigma_{i+1,k^0}\right)(1-H[u_k-\frac{1}{2}\phi_x t])\right] \\
=&\widetilde g_{i+1/2,k}+\widetilde f_{i+1/2,k},
\end{aligned}
\end{equation*}
where $\widetilde g_{i+1/2,k}$ is related to equilibrium state and $\widetilde f_{i+1/2,k}$ is the initial non-equilibrium distribution.

\subsection{Two dimensional case}

The unified gas-kinetic scheme is a multidimensional method,
where both derivatives of flow variables in the normal and tangential directions of a cell interface are taken into account.
With the external force $\vec \phi=\phi_x \vec i+\phi_y \vec j$, the BGK-type model  in the two-dimensional Cartesian coordinate system is
\begin{equation*}
	f_t+uf_x+vf_y+\phi_xf_u+\phi_yf_v=\frac{f^+-f}{\tau},
\end{equation*}
where $\tau=\mu/p$ is the particle collision time and $f^+$ is the  equilibrium distribution.

The integral solution can be written as
\begin{equation}
\begin{aligned}
f(x,y,t,u,v,\xi)=&\frac{1}{\tau}\int_{t^n}^t f^+(x',y',t',u',v',\xi)e^{-(t-t')/\tau}dt' \\
&+e^{-(t-t^n)/\tau}f_0^n(x^n,y^n,t^n,u^n,v^n,\xi),
\end{aligned}
\label{eqn:integral solution 2D}
\end{equation}
where $x'=x-u'(t-t')-\frac{1}{2}\phi_x(t-t')^2, y'=y-v'(t-t')-\frac{1}{2}\phi_y(t-t')^2,u'=u-\phi_x(t-t')$, and $v'=v-\phi_y(t-t')$.

In the unified scheme, at the center of a cell interface $(x_{i+1/2},y_j)$ the solution $f_{i+1/2,j,k,l}$ is constructed from the integral solution Eq.(\ref{eqn:integral solution 2D}).
With the notations $x_{i+1/2}=0,y_j=0$ at $t^n=0$, the time-dependent interface distribution function goes to
\begin{equation*}
\begin{aligned}
f(0,0,t,u_k,v_l,\xi)=&\frac{1}{\tau}\int_{0}^t f^+(x',y',t',u_k',v_l',\xi)e^{-(t-t')/\tau}dt' \\
&+e^{-t/\tau}f_0(-(u_k-\phi_xt)t,-(v_l-\phi_y t)t-\frac{1}{2}\phi_y t^2,0,u_k-\phi_x t,v_l-\phi_y t,\xi),
\end{aligned}
\end{equation*}
where the trajectories are $x'=-u_k'(t-t')-\frac{1}{2}\phi_x(t-t')^2,y'=-v_l'(t-t')-\frac{1}{2}\phi_y(t-t')^2,u_k'=u_k-\phi_x (t-t')$, and $v_l'=v_l-\phi_y (t-t')$.

As a second order scheme, the initial gas distribution function $f_0$ is reconstructed as
\begin{equation*}
f_0(x,y,0,u_k,v_l,\xi)=\left\{
\begin{aligned}
&f_{i+1/2,j,k,l}^L+\sigma_{i,j,k,l}x+\theta_{i,j,k,l}y, \quad x\le 0, \\
&f_{i+1/2,j,k,l}^R+\sigma_{i+1,j,k,l}x+\theta_{i+1,j,k,l}y, \quad x> 0,
\end{aligned}
\right.
\end{equation*}
where $f_{i+1/2,j,k,l}^L$ and $f_{i+1/2,j,k,l}^R$ are the reconstructed initial distribution functions at the left and right hand sides of a cell interface.

The equilibrium distribution function around a cell interface is constructed as
\begin{equation*}
g=g_0\left[1+(1-H[x]){a}^Lx+H[x]{a}^Rx+by+{A}t\right],
\end{equation*}
where $g_0$ is the Maxwellian distribution at $(x=0,t=0)$. Here $a^L,a^R$, and $A$ are from the Taylor expansion of a Maxwellian
\begin{equation*}
\begin{aligned}
&a^{L,R}=a_1^{L,R}+a_2^{L,R}u+a_3^{L,R}v+a_4^{L,R}\frac{1}{2}(u^2+v^2+\xi ^2)=a_{\alpha}^{L,R}\psi_\alpha, \\
&b=b_1+b_2u+b_3v+b_4\frac{1}{2}(u^2+v^2+\xi^2)=b_{\alpha}\psi_\alpha, \\
&A=A_1+A_2u+A_3v+A_4\frac{1}{2}(u^2+v^2+\xi ^2)=A_{\alpha}\psi_\alpha.
\end{aligned}
\end{equation*}

The coefficients above can be determined in the same way as one-dimensional case. The time dependent interface distribution function writes
\begin{equation}
\begin{aligned}
f(0,0,t,u_k,v_l,\xi)=&\left(1-e^{-t/\tau}\right)(g_0'+g'^+)\\
&+\left(\tau(-1+e^{-t/\tau})+te^{-t/\tau}\right)a^{L,R}u_kg_0' \\
&-\left[\tau\left(\tau(-1+e^{-t/\tau})+te^{-t/\tau}\right)+\frac{1}{2}t^2e^{-t/\tau}\right] a^{L,R}\phi_x g_0' \\
&+\left(\tau(-1+e^{-t/\tau})+te^{-t/\tau}\right)bv_lg_0' - \left[\tau\left(\tau(-1+e^{-t/\tau})+te^{-t/\tau}\right)+\frac{1}{2}t^2e^{-t/\tau}\right] b\phi_y g_0' \\
&+\tau \left(t/\tau-1+e^{-t/\tau}\right){A}g_0 \\
&+e^{-t/\tau}\left[\left(f_{i+1/2,k^0,l^0}^L+\left(-(u_k-\phi_xt)t-\frac{1}{2}\phi_xt^2\right)\sigma_{i,k^0,l^0}\right.\right.\\
&\left.\left. +\left(-(v_l-\phi_yt)t-\frac{1}{2}\phi_yt^2\right)\theta_{i,k^0,l^0}\right)H[u_k-\frac{1}{2}\phi_x t] \right.\\
&\left.+\left(f_{i+1/2,k^0,l^0}^R+\left(-(u_k-\phi_xt)t-\frac{1}{2}\phi_xt^2\right)\sigma_{i+1,k^0,l^0}\right. \right. \\
&\left. \left. +\left(-(v_l-\phi_yt)t-\frac{1}{2}\phi_yt^2\right)\theta_{i+1,k^0,l^0}\right)(1-H[u_k-\frac{1}{2}\phi_x t])\right] \\
=&\widetilde g_{i+1/2,j,k,l}+\widetilde f_{i+1/2,j,k,l},
\end{aligned}
\end{equation}
where $\widetilde g_{i+1/2,j,k,l}$ is related to equilibrium state integration and $\widetilde f_{i+1/2,j,k,l}$ is the initial non-equilibrium distribution.
The extension of the above method to three dimensional case can be done similarly.

\subsection{Update algorithm}
With the cell averaged distribution function
\begin{equation*}
f_{x_i,y_j,t^n,u_k,v_l}=f_{i,j,k,l}^n=\frac{1}{\Omega_{i,j}(\vec x)\Omega_{k,l}(\vec u)} \int_{\Omega_{i,j}} \int_{\Omega_{k,l}}f(x,y,t^n,u,v)d\vec x d\vec u,
\end{equation*}
the direct modeling for the flow evolution in a discretized space gives
\begin{equation}
\begin{aligned}
f_{i,j,k,l}^{n+1}=&f_{i,j,k,l}^n+\frac{1}{\Omega_{i,j}}\int_{t^n}^{t^{n+1}} \sum_{r=1} u_r\hat f_r(t)\Delta S_r dt\\
&+\frac{1}{\Omega_{i,j}}\int_{t^n}^{t^{n+1}}\int_{\Omega_{i,j}}Q(f)d\vec x dt+\frac{1}{\Omega_{i,j}}\int_{t^n}^{t^{n+1}}\int_{\Omega_{i,j}}G(f)d\vec x dt,
\end{aligned}
\label{eqn:distribution update}
\end{equation}
where $\hat f_r$ is the time-dependent gas distribution function at cell interface. $Q(f)$ and $G(f)$ are the source term from collision term and gravitational field,
\begin{equation*}
\begin{aligned}
&Q(f)=\frac{f_{i,j,k,l}^{+}-f_{i,j,k,l}^{n+1/2}}{\tau}, \\
&G(f)=-\phi_x \frac{\partial}{\partial u} f_{i,j,k,l}^{n+1/2}-\phi_y \frac{\partial}{\partial v} f_{i,j,k,l}^{n+1/2}.
\end{aligned}
\end{equation*}

In the UGKS, we use the semi-implicit method to model the source term of distribution function
\begin{equation}
\begin{aligned}
f_{i,j,k,l}^{n+1}=& f_{i,j,k,l}^n+\frac{1}{\Omega_{i,j}}\left(F_{i-1/2,j,k,l}-F_{i+1/2,j,k,l}\right)+\frac{1}{\Omega_{i,j}}\left(F_{i,j-1/2,k,l}-F_{i,j+1/2,k,l}\right)  \\
& +\frac{\Delta t}{2}\left(\frac{f_{i,j,k,l}^{+(n+1)}-f_{i,j,k,l}^{n+1}}{\tau^{n+1}}+\frac{f_{i,j,k,l}^{+(n)}-f_{i,j,k,l}^n}{\tau^{n}}\right)-\phi_x \Delta t \frac{\partial}{\partial u} f_{i,j,k,l}^{n+1}-\phi_y \Delta t\frac{\partial}{\partial v} f_{i,j,k,l}^{n+1},
\end{aligned}
\label{eqn:micro update}
\end{equation}
where the derivatives of particle velocity are evaluated via implicit upwind finite difference method in the discretized velocity space.

In order to update the gas distribution function, let's take  conservative moments on Eq.(\ref{eqn:distribution update}) first.
The updates of the conservation flow variables are
\begin{equation}
\textbf{W}_{i,j}^{n+1}=\textbf{W}_{i,j}^n+\frac{1}{\Omega_{i,j}}\int_{t^n}^{t^{n+1}}\sum_{r=1}\Delta \textbf{S}_r\cdot {\textbf{F}}_rdt+\frac{1}{\Omega_{i,j}}\int_{t^n}^{t^{n+1}}\textbf{G}_{i,j}dt
\label{eqn:macro update}
\end{equation}
where $\textbf{F}_r$ are the fluxes of conservative flow variables, and $\textbf{G}_{i,j}$ is the source term from external force,
\begin{equation}
\textbf{G}_{i,j}=\int_{\Omega_{k,l}}\left(-\phi_x \Delta t \frac{\partial}{\partial u} f_{i,j,k,l}-\phi_y \Delta t\frac{\partial}{\partial v} f_{i,j,k,l}\right)\psi dudvd\xi.
\end{equation}
Eq.(\ref{eqn:macro update}) can be solved first, and its solution can be used for the construction of the equilibrium state in Eq.(\ref{eqn:micro update}) at $t^{n+1}$.
Subsequently, the implicit Eq.(\ref{eqn:micro update}) for $f^{n+1}$ can be solved explicitly.

\section{Numerical experiments}

In this section, we are going to present numerical examples to validate the well-balanced UGKS.
In order to demonstrate the  capability of the scheme to resolve multi-scale flow physics, simulations from free molecule flow to continuum Euler and NS solutions under the gravitational field will be presented.
The flow features in different regimes can be well captured by the unified scheme.
For the first time, an interesting non-equilibrium phenomena in the lid-driven cavity case, such as the correlation between the heat flux and the gravitational field,
will be demonstrated. The Shakhov model is used for the construction of UGKS, and hard sphere (HS) monatomic perfect gas is employed in all test cases.

\subsection{One-dimensional hydrostatic equilibrium solution}
\begin{figure}[htb!]
	\centering
	\includegraphics[width=6cm]{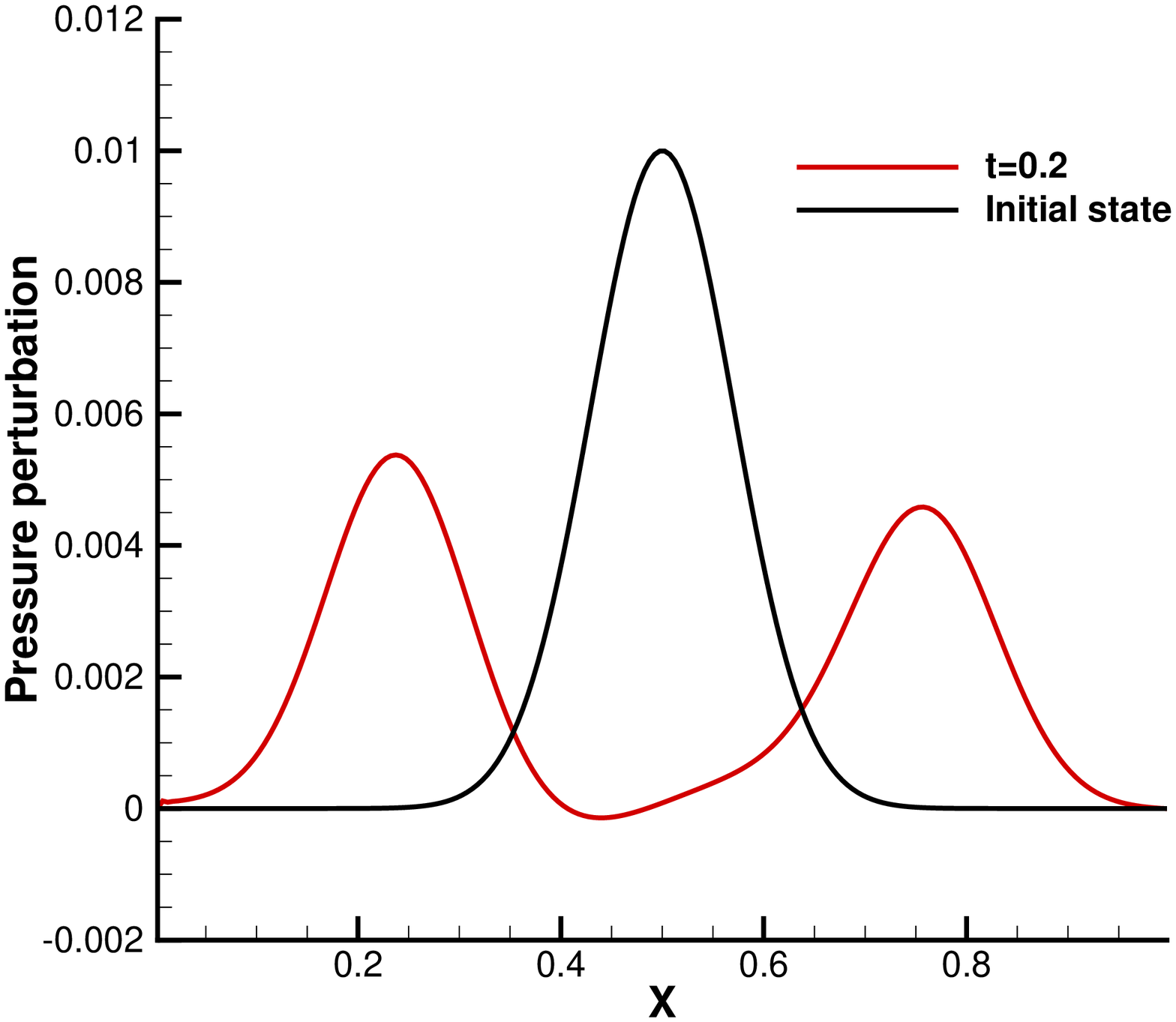}
	\caption{Pressure perturbation of the hydrostatic equilibrium solution}
	\label{pic:perturbation}
\end{figure}

The first case originates from Leveque and Bale's paper \cite{leveque1999wave}. In the simulation, a monatomic ideal gas with $\gamma=5/3$ is initially set up with a hydrostatic equilibrium state in the domain $x\in [0,1]$ under the gravitational field $\phi_x=-1.0$ pointing towards to the negative $x$-direction,
\begin{equation*}
\rho_0(x)=p_0(x)=\exp(-x), u_0(x)=0.
\end{equation*}
The test is for the solution with an added instant perturbation to the initial pressure,
\begin{equation*}
p(x,t=0)=p_0(x)+\eta\exp(-100(x-5.0)^2),
\end{equation*}
where $\eta=0.01$ is a constant.
The computational domain is divided into 200 uniform cells and velocity space with $100$ discretized velocity points.
The Knudsen number for this test case has a value $10^{-4}$, which is basically in the continuum flow regime.
The simulation result at $t=0.2$  is presented in Fig. \ref{pic:perturbation}. As analyzed in \cite{leveque1999wave,tian2007three},
operator splitting methods fail to capture such a small perturbation solution,
and the effect of gravity should be explicitly considered in the flux evaluation for a well-balanced scheme.
It is clear that exact hydrostatic solution under gravity is well preserved by UGKS during the spreading process of the perturbation.

\subsection{Shock tube problem under gravitational field}

The second case is the standard Sod shock tube problem under gravitational field \cite{xuluochen,luo2011well}.
The computational domain is $x\in [0,1]$, which is divided into 100 cells. The simulation uses monatomic gas with $\gamma=5/3$ and non-reflection boundary condition at both ends.

The initial condition is set as
\begin{equation*}
\rho=1.0,U=0.0,p=1.0,  x\le 0.5,
\end{equation*}
\begin{equation*}
\rho=0.125,U=0.0,p=0.1,  x>0.5.
\end{equation*}

The gravity $\phi_x=-1.0$ is in the opposite direction of $x$ axis.
The simulation results at $t=0.2$ are presented. In order to present the capability of the unified scheme to simulate flow physics in different flow regimes,
we perform the simulations with different reference Knudsen number, such as $\rm Kn=0.0001$, $\rm Kn=0.01$ and $\rm Kn=1$,
which correspond to typical continuum, transition, and free molecular transport. The reference Knudsen number is used to define dynamic viscosity in the reference state via variable soft sphere model (VSS),
\begin{equation}
\mu_{ref}=\frac{5(\alpha+1)(\alpha+2)\sqrt{\pi}}{4\alpha(5-2\omega)(7-2\omega)}Kn_{ref}.
\label{eqn: vss viscosity}
\end{equation}
In this simulation, we choose $\alpha=1.0$ and $\omega=0.5$ to recover a hard sphere monatomic gas.
The viscosity for the hard-sphere model is,
\begin{equation}
\mu=\mu_{ref}\left(\frac{T}{T_{ref}}\right)^\theta,
\label{eqn: hs model}
\end{equation}
where $T_{ref}$ is the reference temperature and $\theta$ is the index related to HS model. In this case we adopt the value $\theta=0.72$. The local collision time is evaluated with the relation $\tau=\mu/p$.

\begin{figure}[htb!]
	\centering
	\subfigure[Density]{
		\includegraphics[width=6cm]{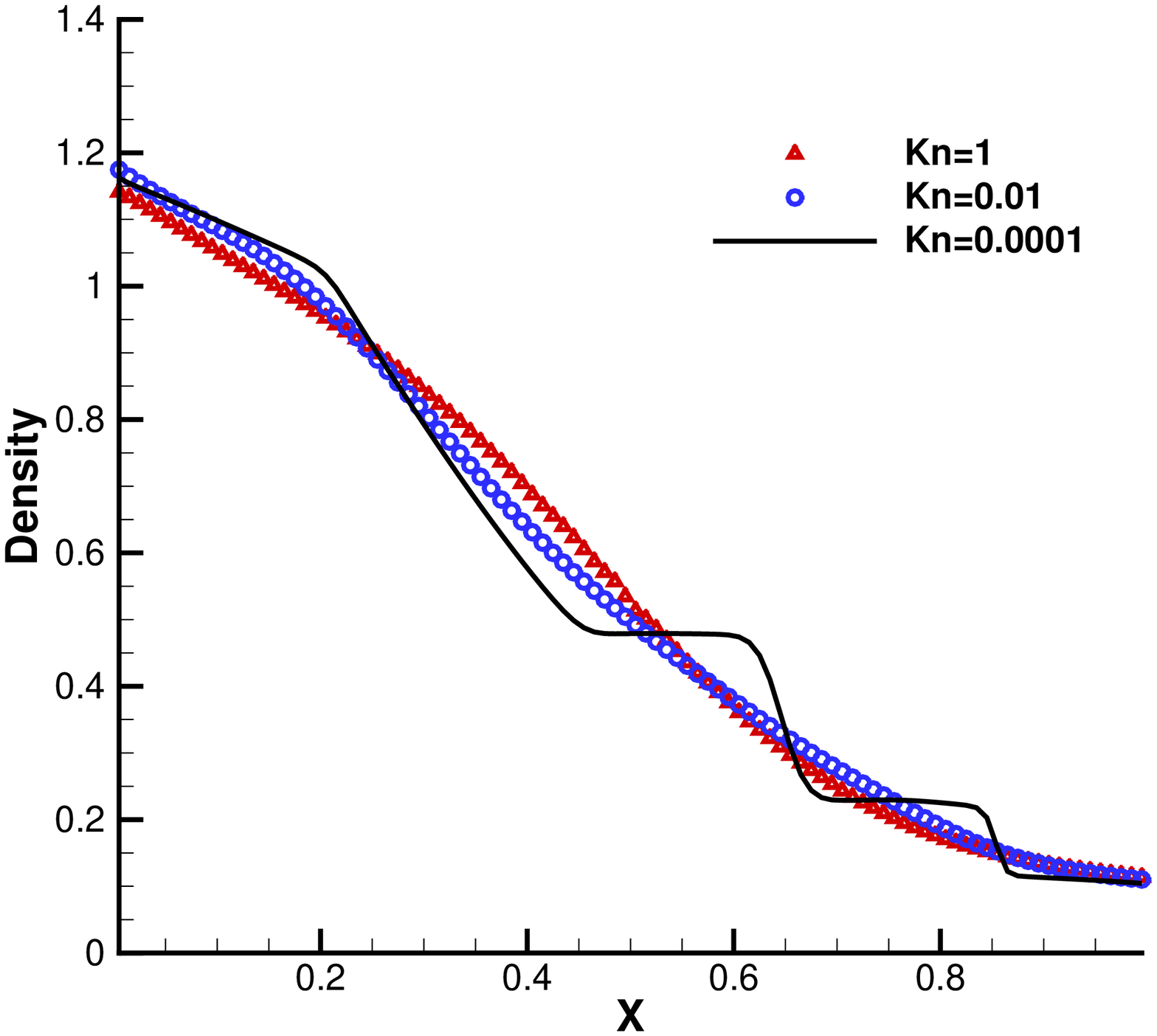}
	}
	\subfigure[Velocity]{
		\includegraphics[width=6cm]{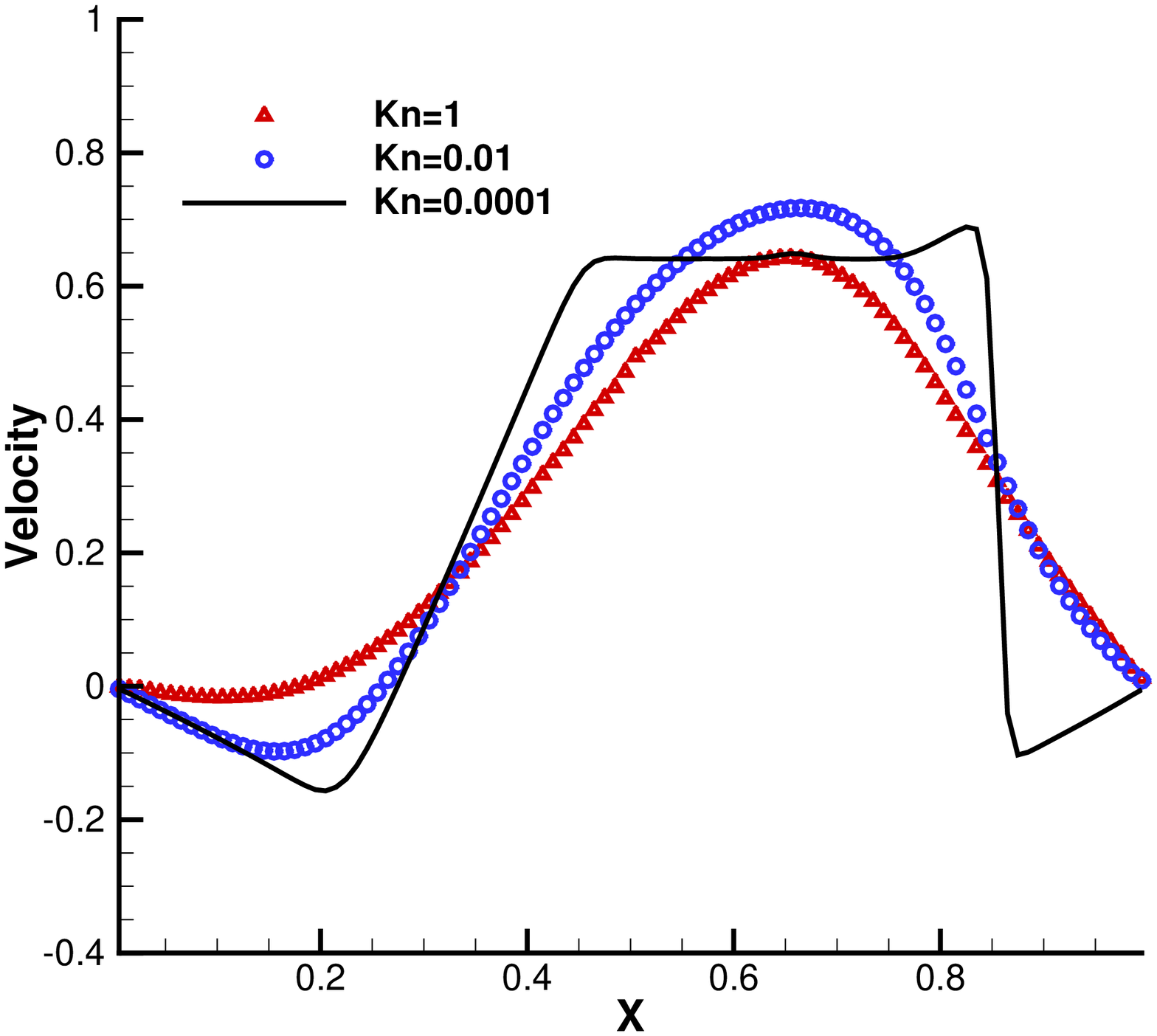}
	}
	\subfigure[Temperature]{
		\includegraphics[width=6cm]{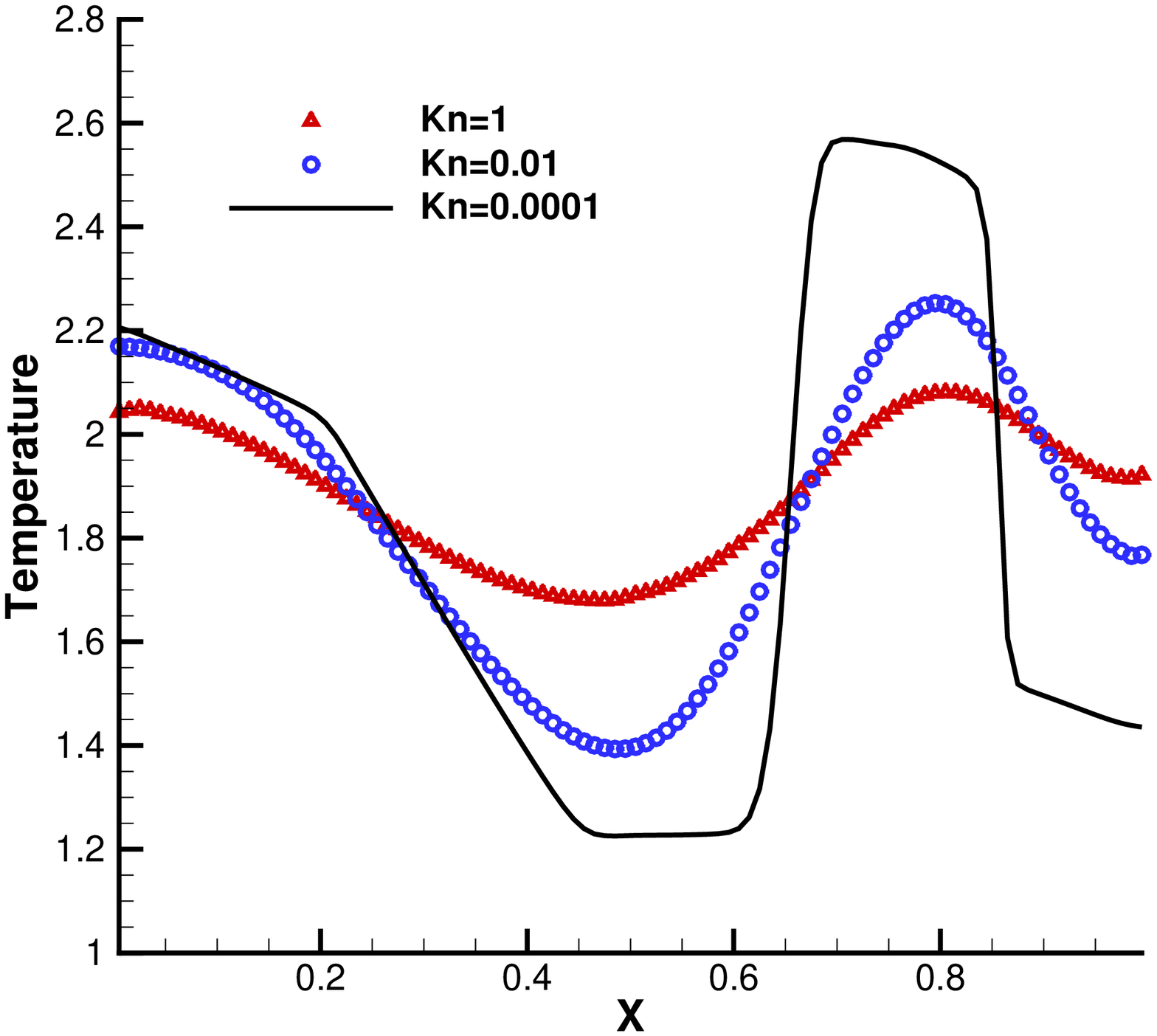}
	}
	\subfigure[Pressure]{
		\includegraphics[width=6cm]{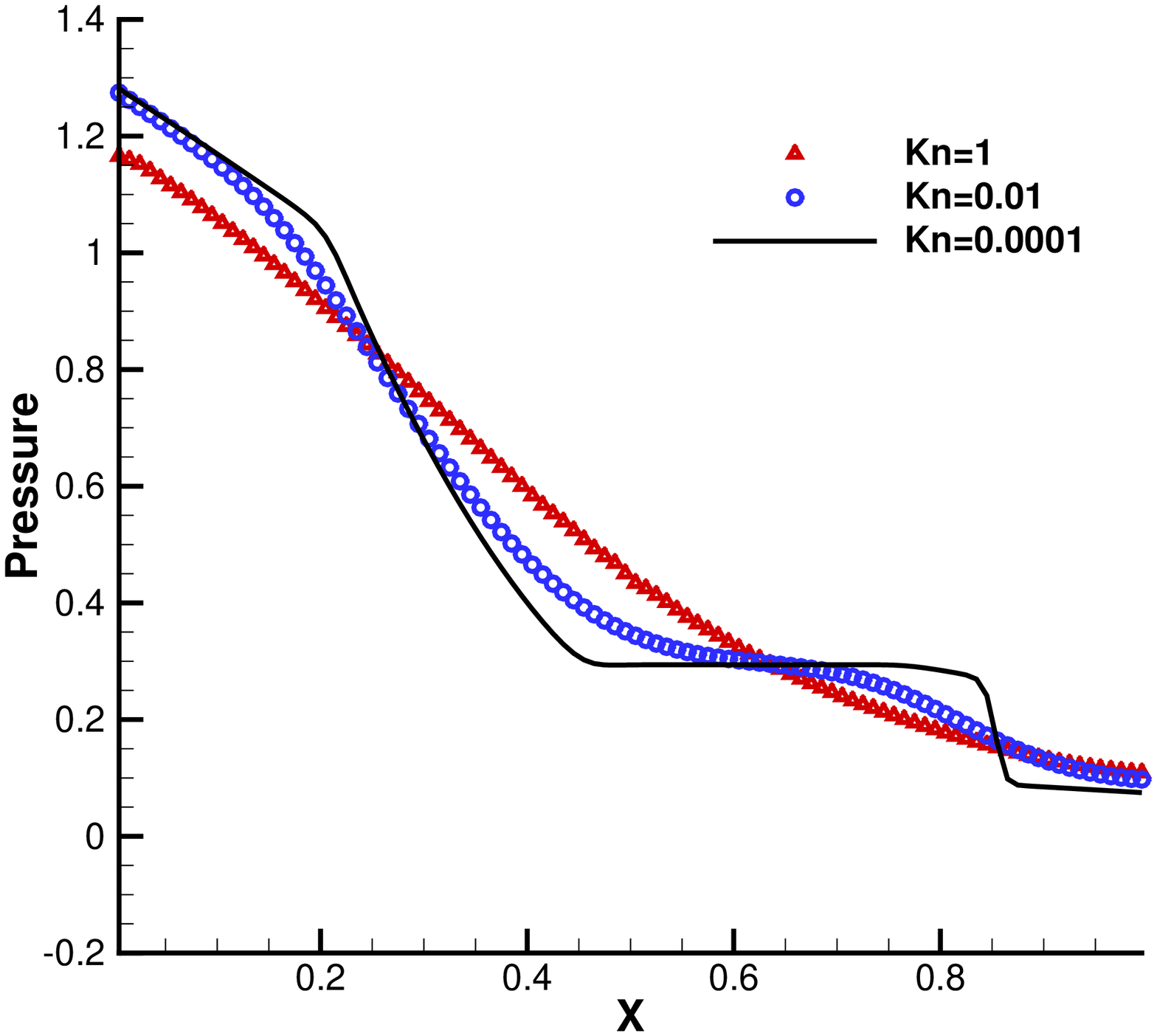}
	}
	\caption{Sod test under gravity.}
	\label{pic:sod}
\end{figure}

The computational results are presented in Fig. \ref{pic:sod}.
It can be observed that under the gravitational potential, the particles inside the tube are "pulled back" in the negative $x$-direction.
In comparison with the case without gravity, such as the standard Sod text case,
the particle moves towards the left hand side of the tube, which results in rising the density, temperature, pressure, and negative flow velocity in some region.

This test case illustrates the capacity of the unified scheme to simulate flow physics in different regimes under gravitational field.
In the continuum regime with $\rm Kn=0.0001$, the collision time is much less than the time step, which results in the Euler solution
and the unified scheme becomes a shock capturing scheme due to the limited resolution in space and time.
With the increment of Knudsen number, the collision time increases and the flow physics changes as well.
There is a smooth transion from the Euler solution of the Riemann problem to collisionless Boltzmann solution.

\subsection{Rayleigh-Taylor instability}

This test case comes from \cite{leveque1999wave}. Consider an isothermal static ideal gas with density and pressure satisfying two segmented exponential relations in a two-dimensional polar coordinate $(r,\theta)$,
\begin{equation*}
\rho_0(r)=e^{-\alpha(r+r_0)},p_0(r)=\frac{1.5}{\alpha}e^{-\alpha(r+r_0)},U_0=0,
\end{equation*}
where
\begin{equation*}
\left\{
\begin{aligned}
&\alpha=2.68,r_0=0.258,\ r\le r_1, \\
&\alpha=5.53,r_0=-0.308,\ r> r_1,
\end{aligned}
\right.
\rm{and}
\left\{
\begin{aligned}
&r_1=0.6(1+0.02\cos(20\theta)), \ \rm{for \  density}, \\
&r_1=0.62324965, \ \rm{for \ pressure}.
\end{aligned}
\right.
\end{equation*}

The external force potential satisfies $d\Phi/dr=1.5$, resulting in a force pointing towards the coordinate origin.
The initial condition contains a density inversion in the flow region,
so there will be an Rayleigh-Taylor instability around the interface under the gravitational field.
The fluid motion around the Rayleigh-Taylor unstable interface is expected to be well resolved by the numerical scheme.
At the same time, a well-balanced scheme should be able to keep the hydrostatic solution away from the interface.

\begin{figure}[htb!]
	\centering
	\subfigure[t=0]{
		\includegraphics[width=3.5cm]{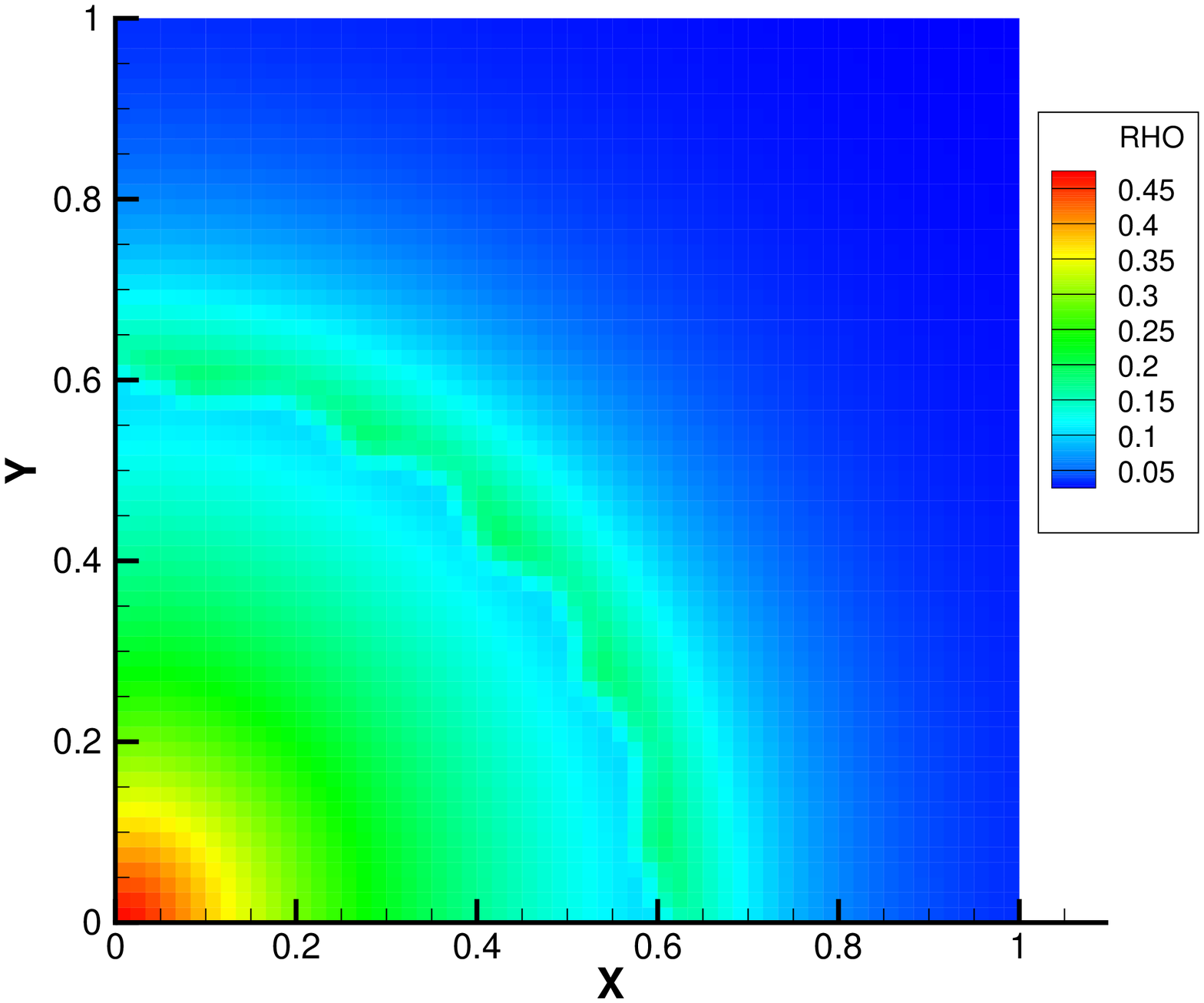}
	}
	\subfigure[t=0.8]{
		\includegraphics[width=3.5cm]{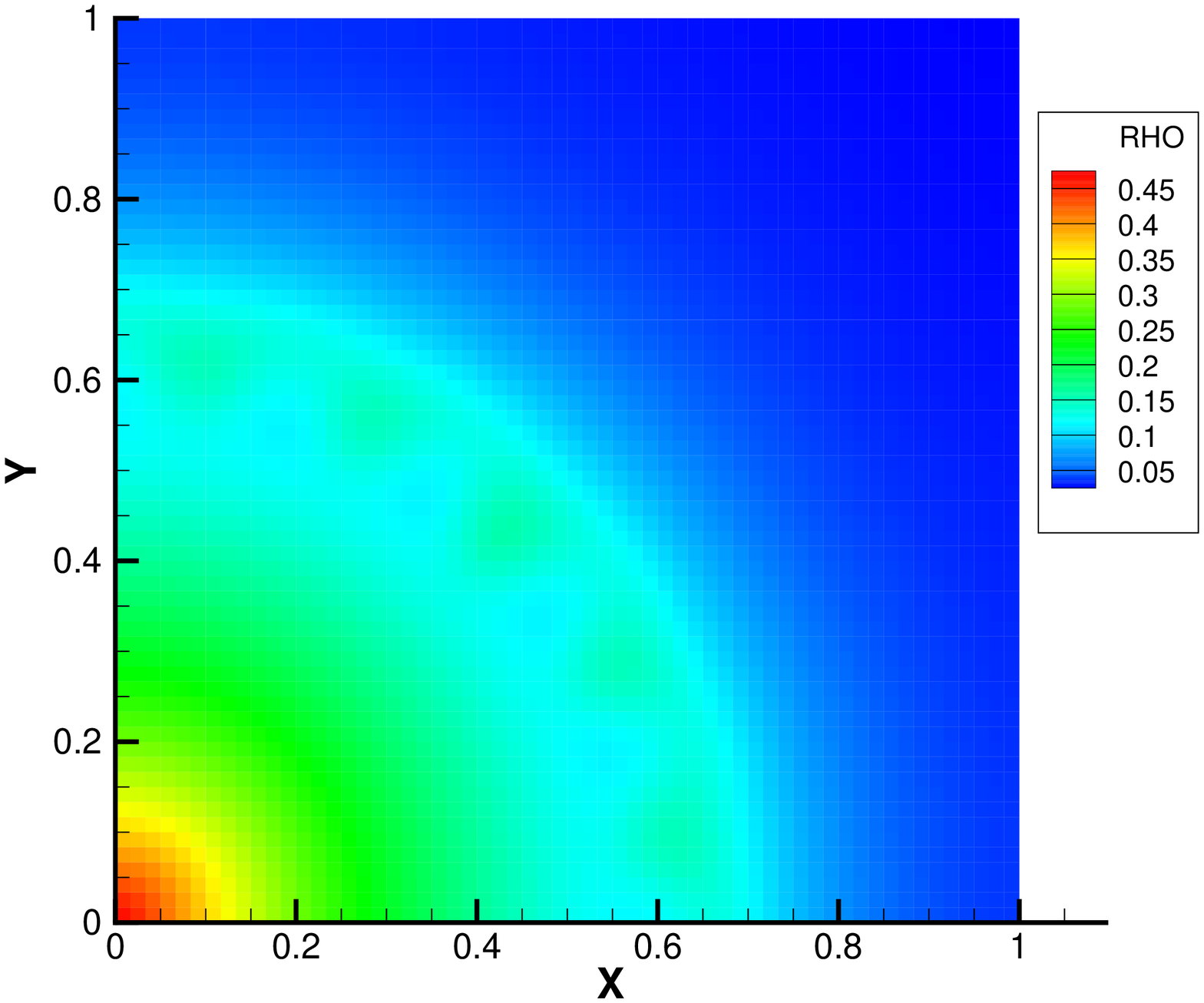}
	}
	\subfigure[t=1.2]{
		\includegraphics[width=3.5cm]{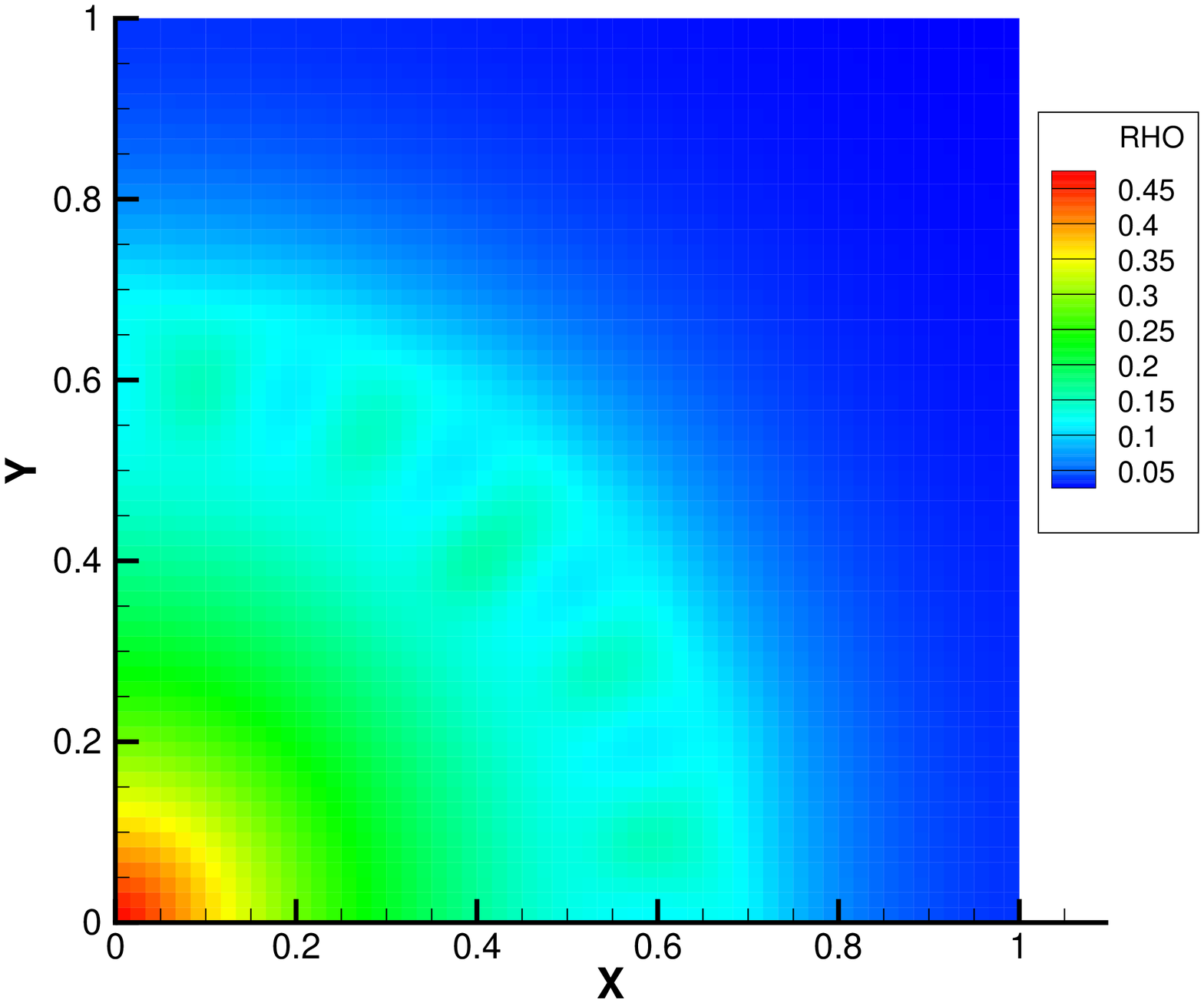}
	}
	\subfigure[t=2.0]{
		\includegraphics[width=3.5cm]{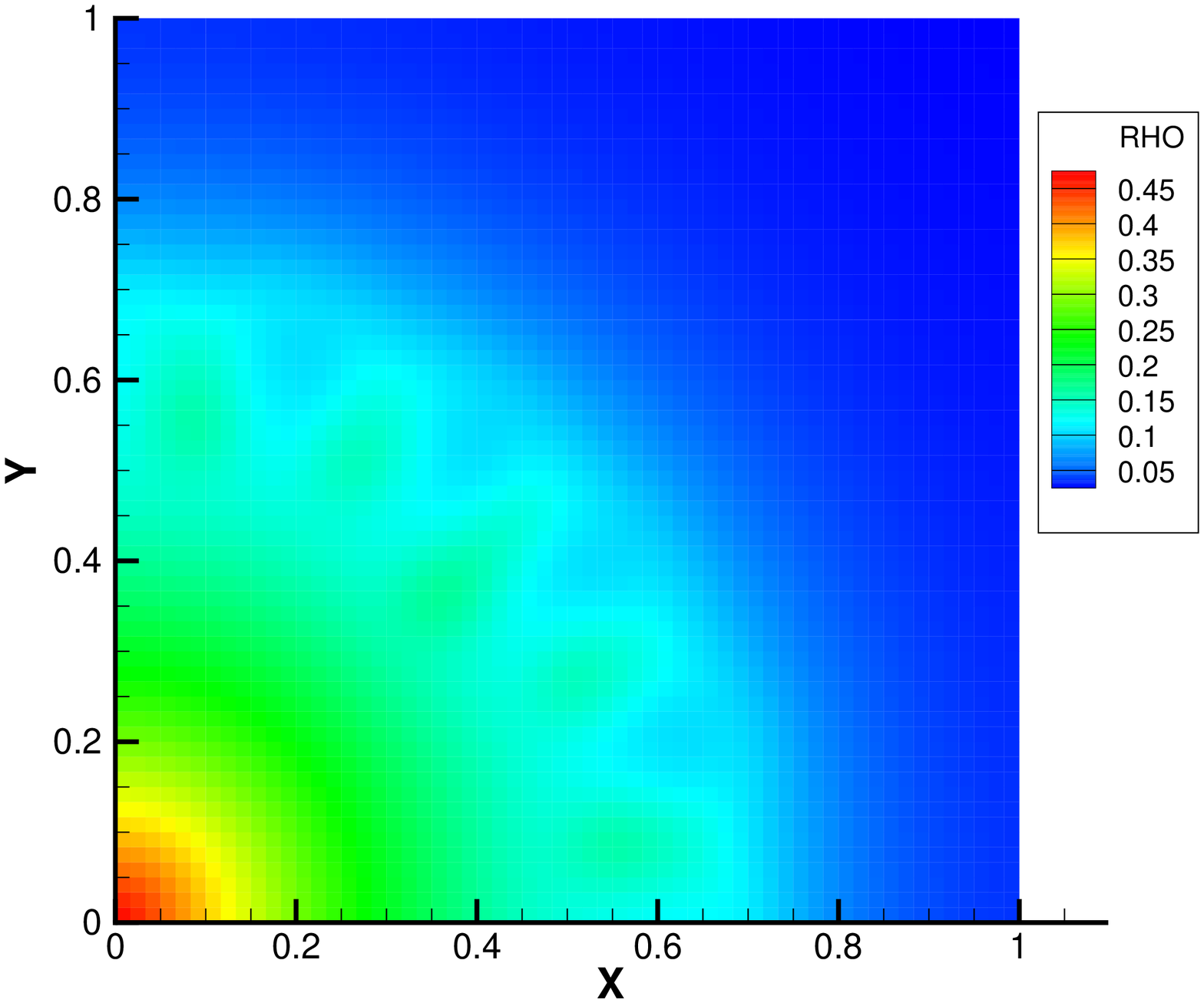}
	}
	\subfigure[t=0]{
		\includegraphics[width=3.5cm]{rtinitialcontour.eps}
	}
	\subfigure[t=0.08]{
		\includegraphics[width=3.5cm]{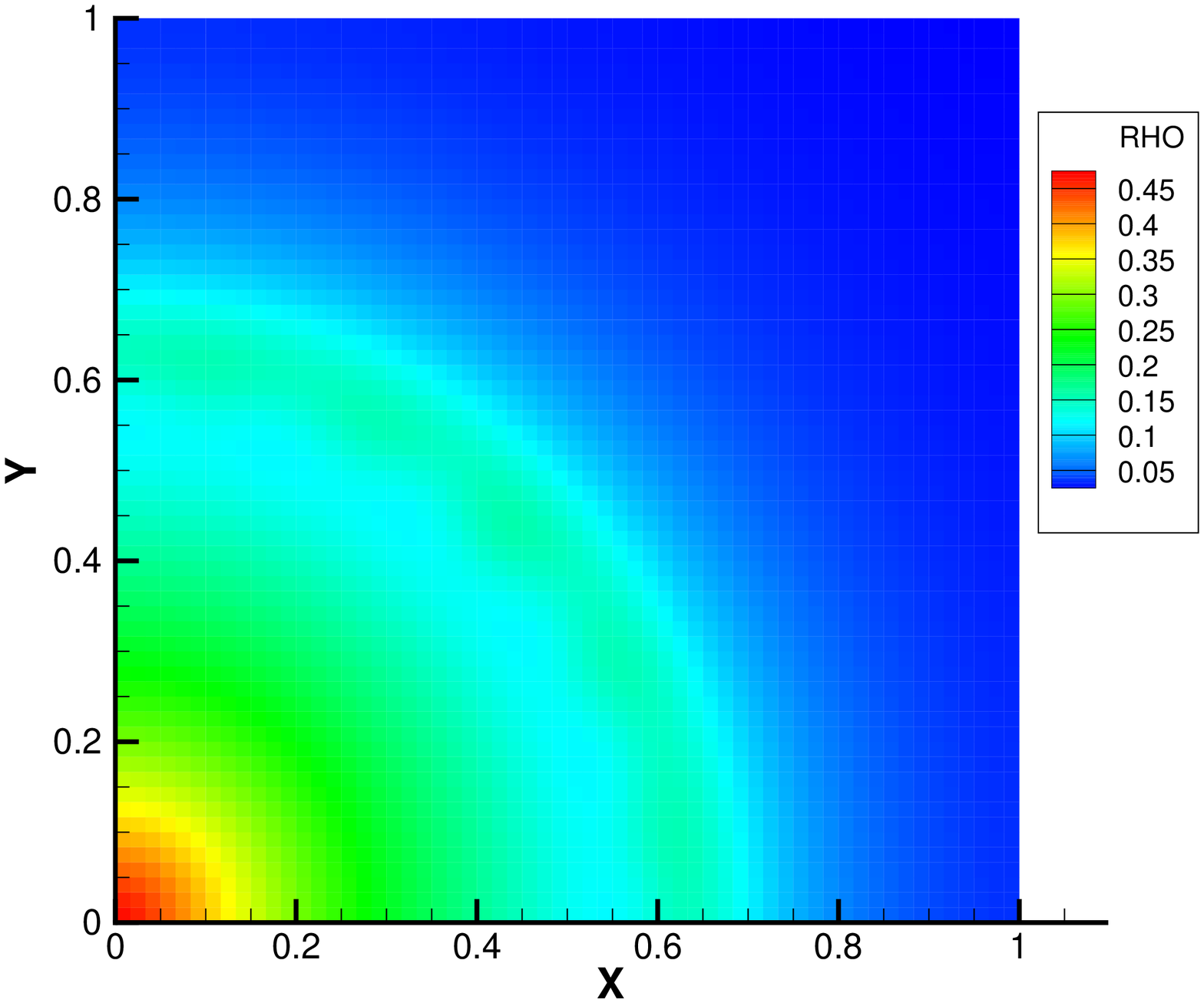}
	}
	\subfigure[t=0.16]{
		\includegraphics[width=3.5cm]{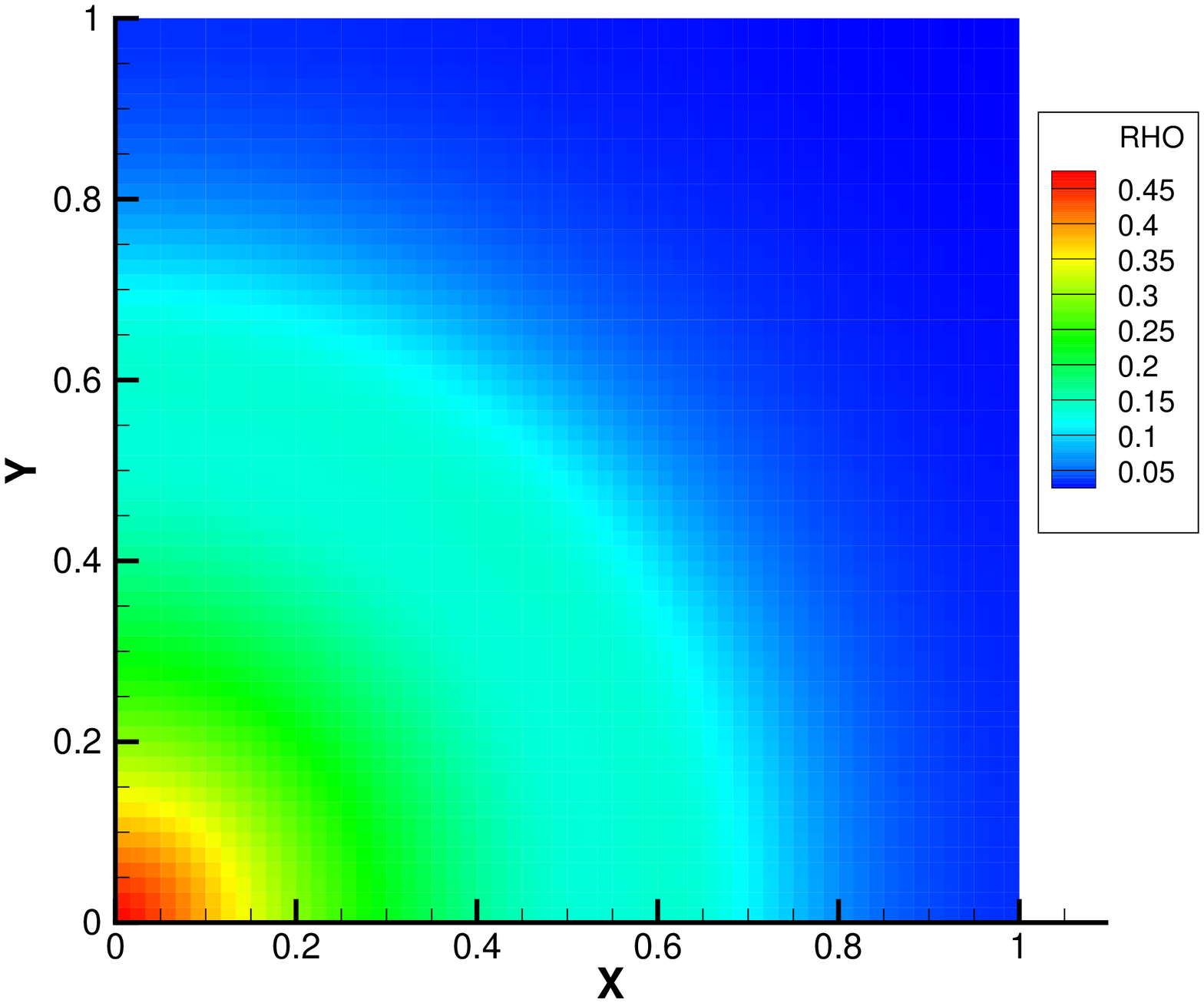}
	}
	\subfigure[t=0.24]{
		\includegraphics[width=3.5cm]{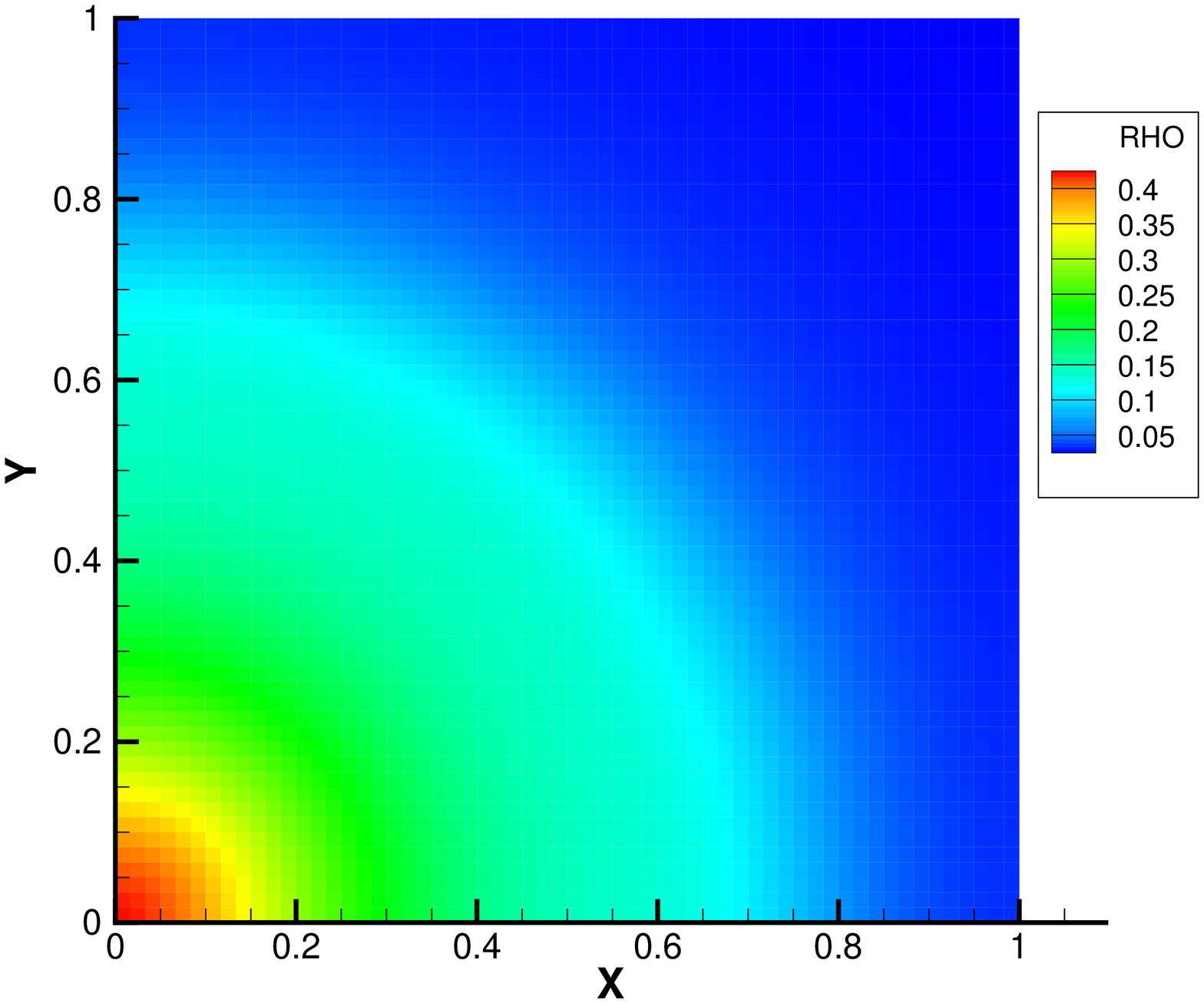}
	}
	\subfigure[t=0]{
		\includegraphics[width=3.5cm]{rtinitialcontour.eps}
	}
	\subfigure[t=0.08]{
		\includegraphics[width=3.5cm]{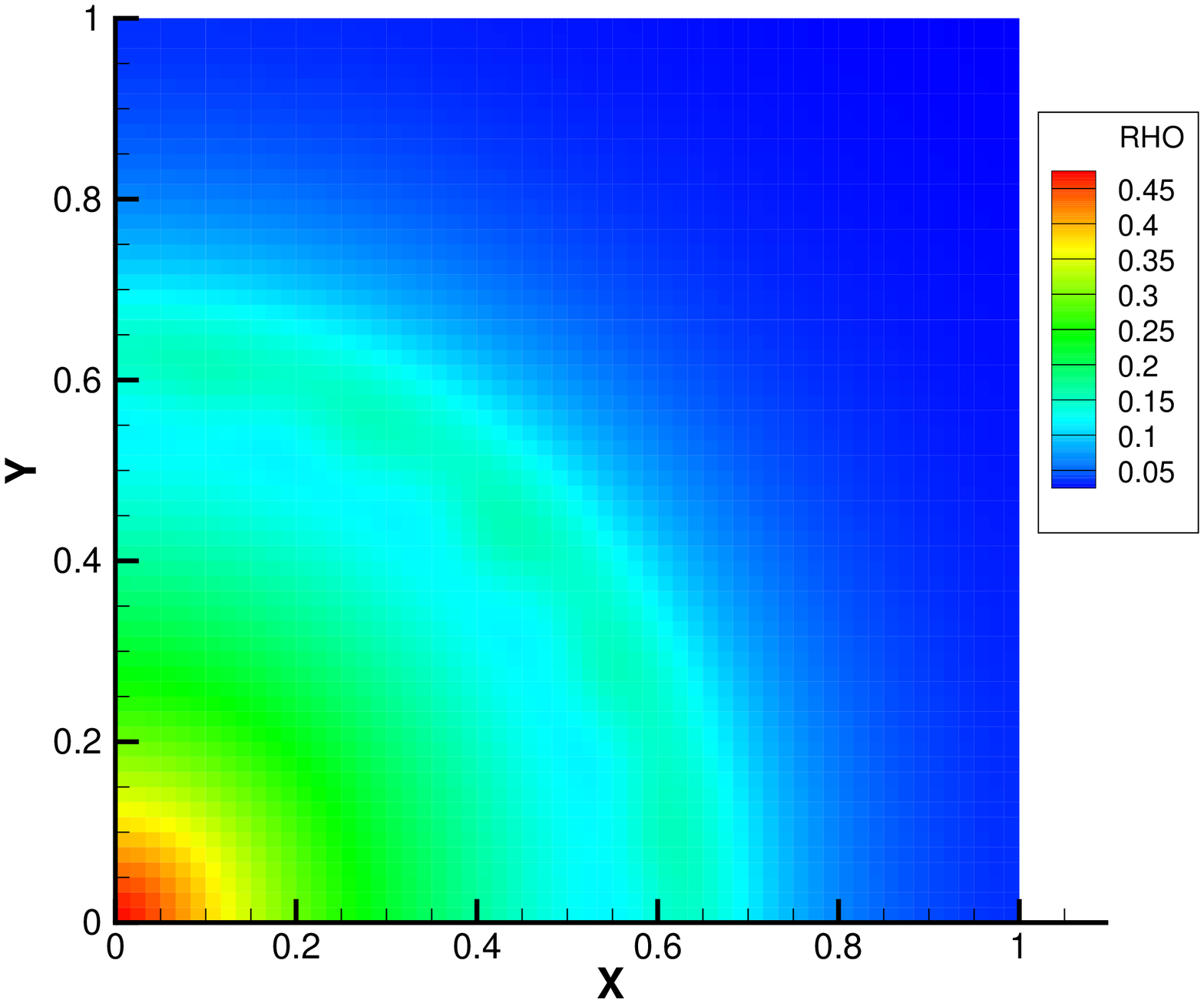}
	}
	\subfigure[t=0.16]{
		\includegraphics[width=3.5cm]{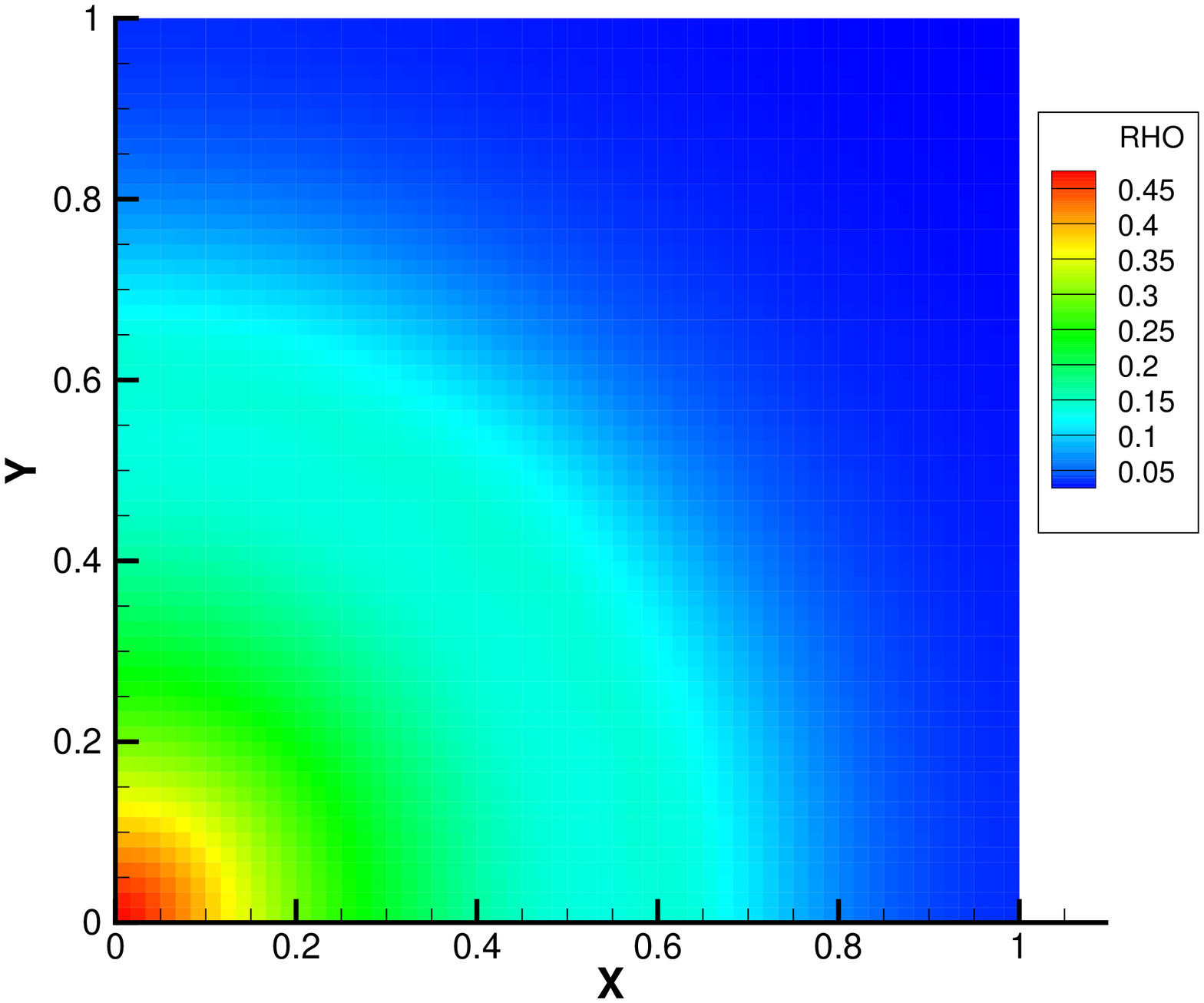}
	}
	\subfigure[t=0.24]{
		\includegraphics[width=3.5cm]{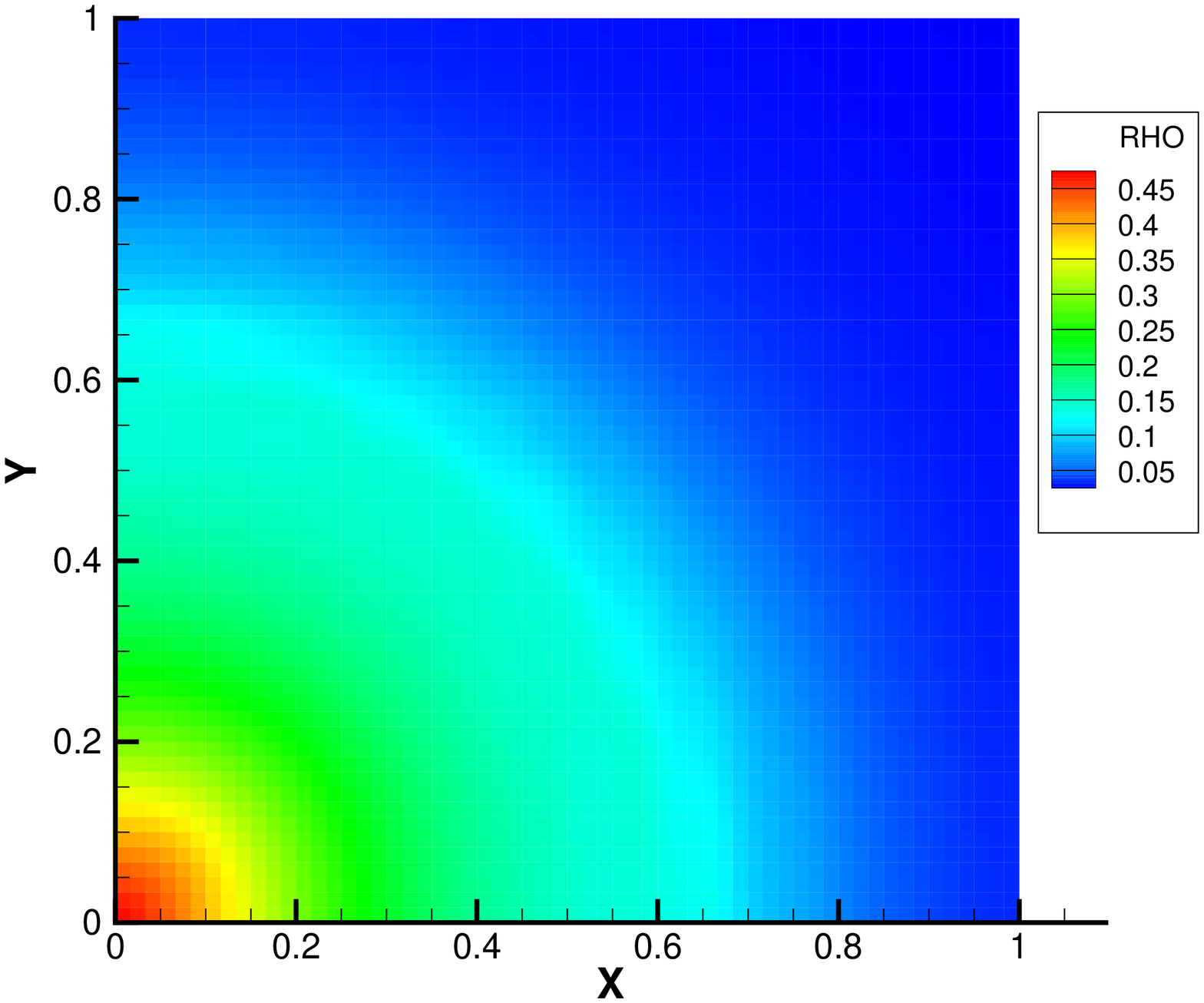}
	}
	\caption{Density evolution under gravity with reference Knudsen number 0.0001, 0.01, 1}
	\label{pic:rt contour}
\end{figure}

\begin{figure}[htb!]
	\centering
	\subfigure[t=0]{
		\includegraphics[width=3.5cm]{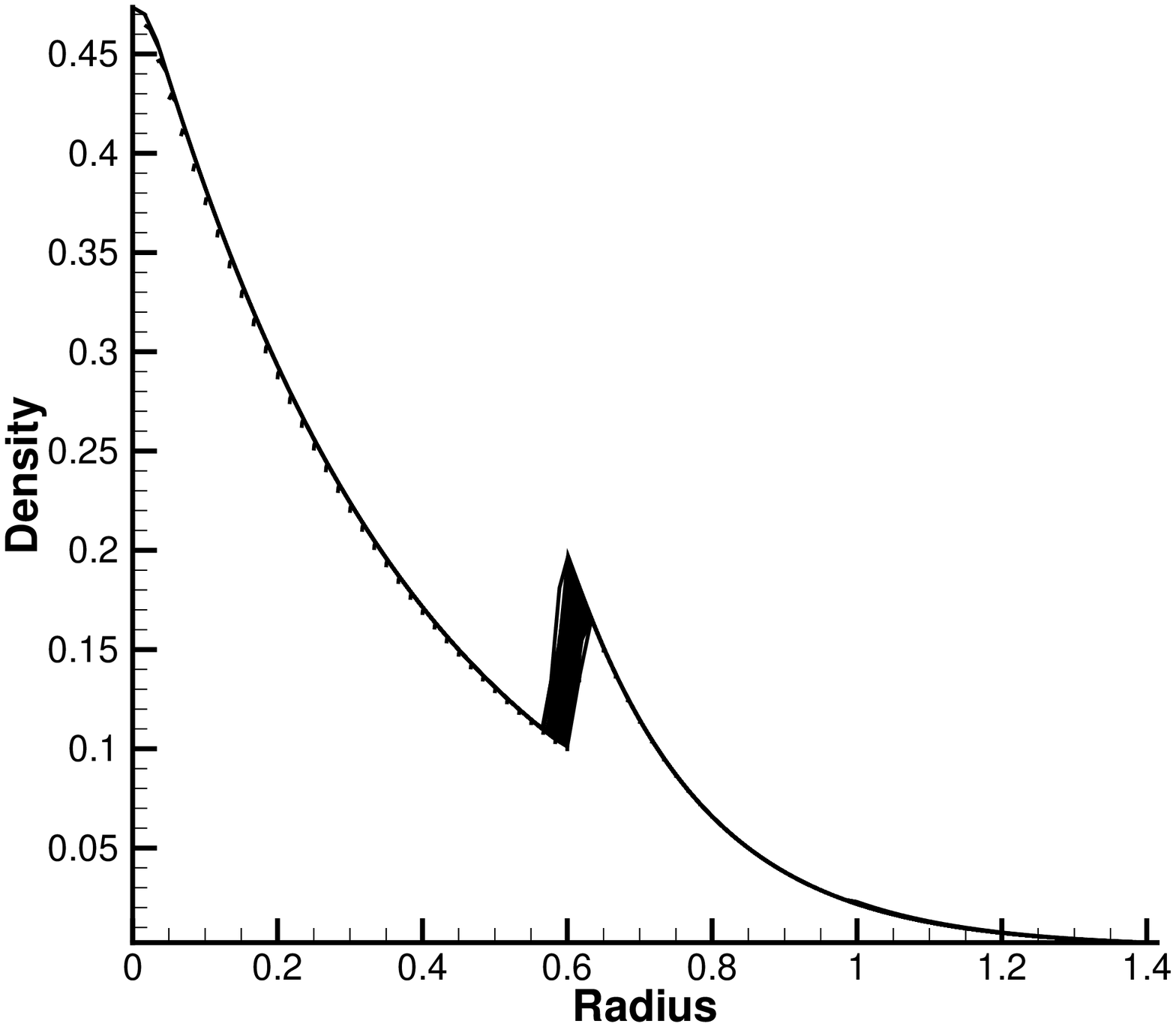}
	}
	\subfigure[t=0.8]{
		\includegraphics[width=3.5cm]{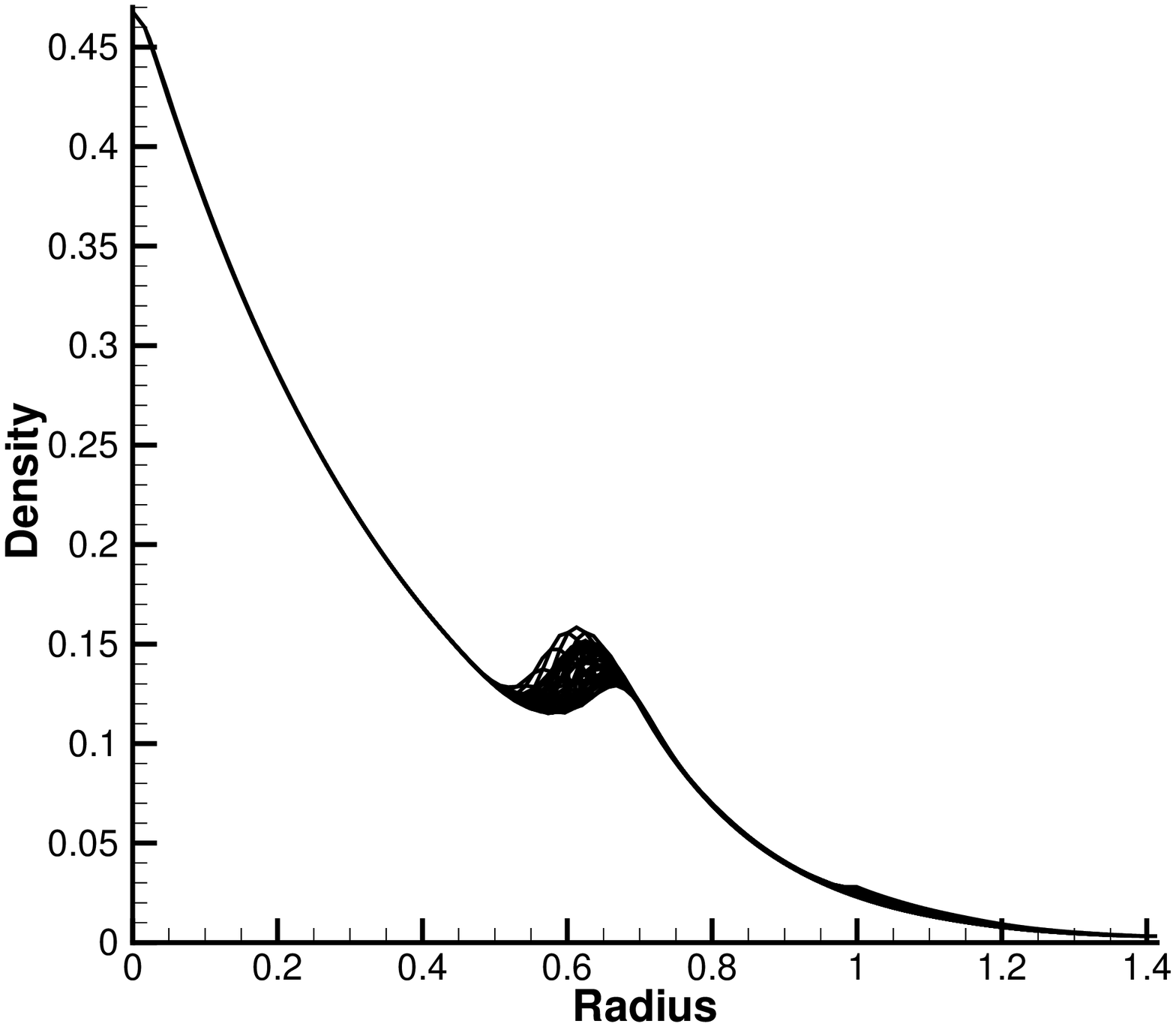}
	}
	\subfigure[t=1.4]{
		\includegraphics[width=3.5cm]{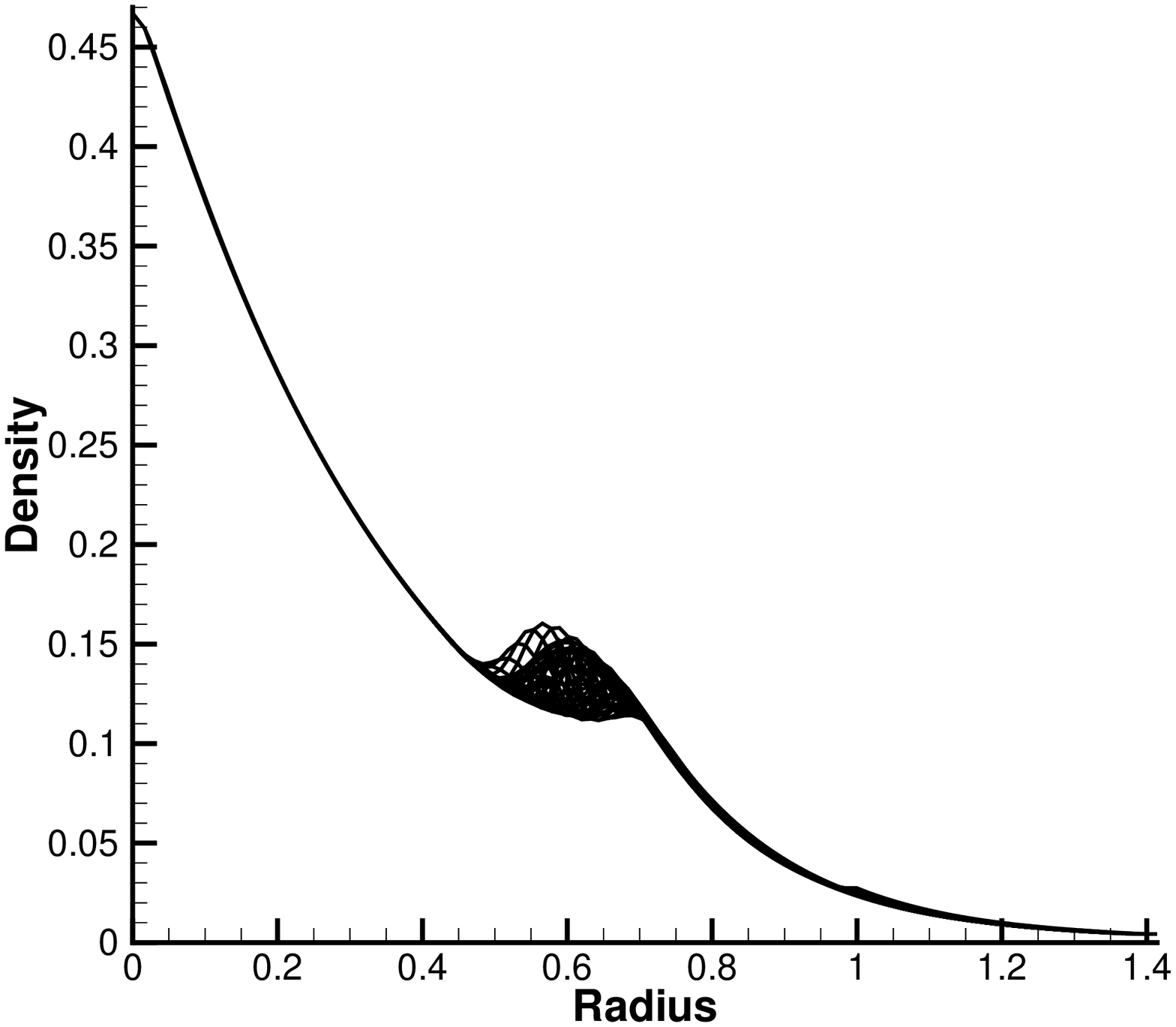}
	}
	\subfigure[t=2.0]{
		\includegraphics[width=3.5cm]{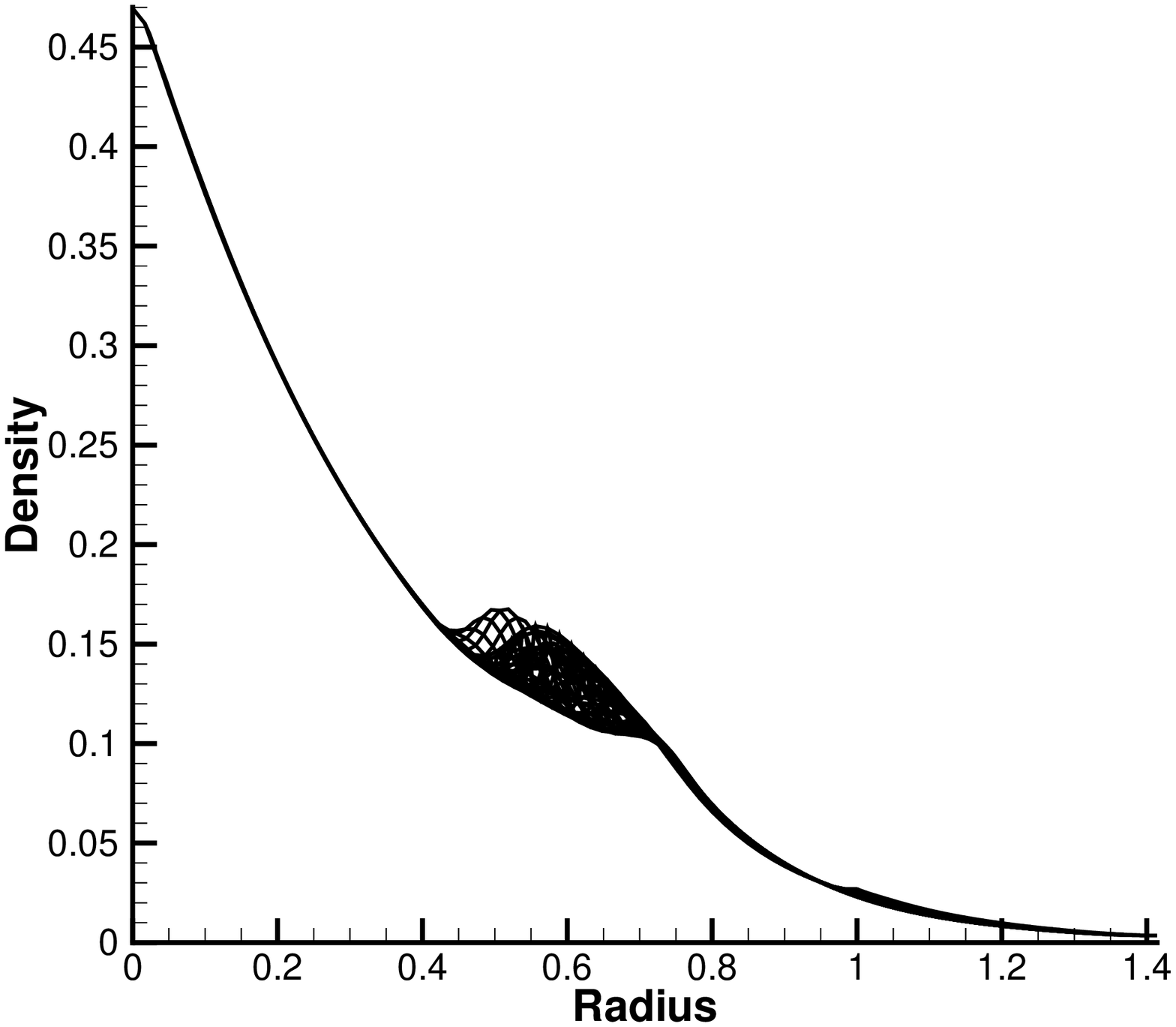}
	}
	\subfigure[t=0]{
		\includegraphics[width=3.5cm]{rtinitialline.eps}
	}
	\subfigure[t=0.08]{
		\includegraphics[width=3.5cm]{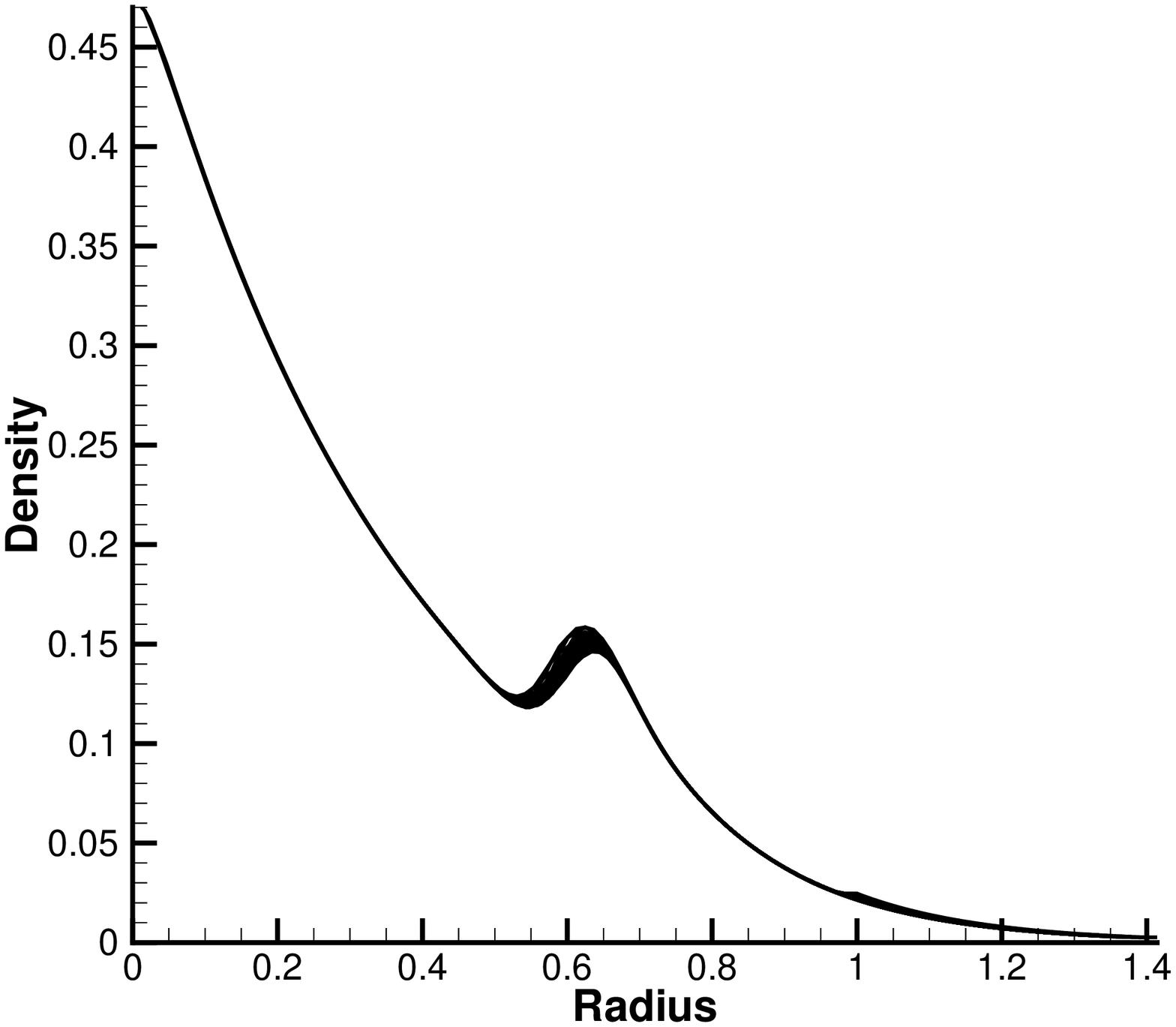}
	}
	\subfigure[t=0.16]{
		\includegraphics[width=3.5cm]{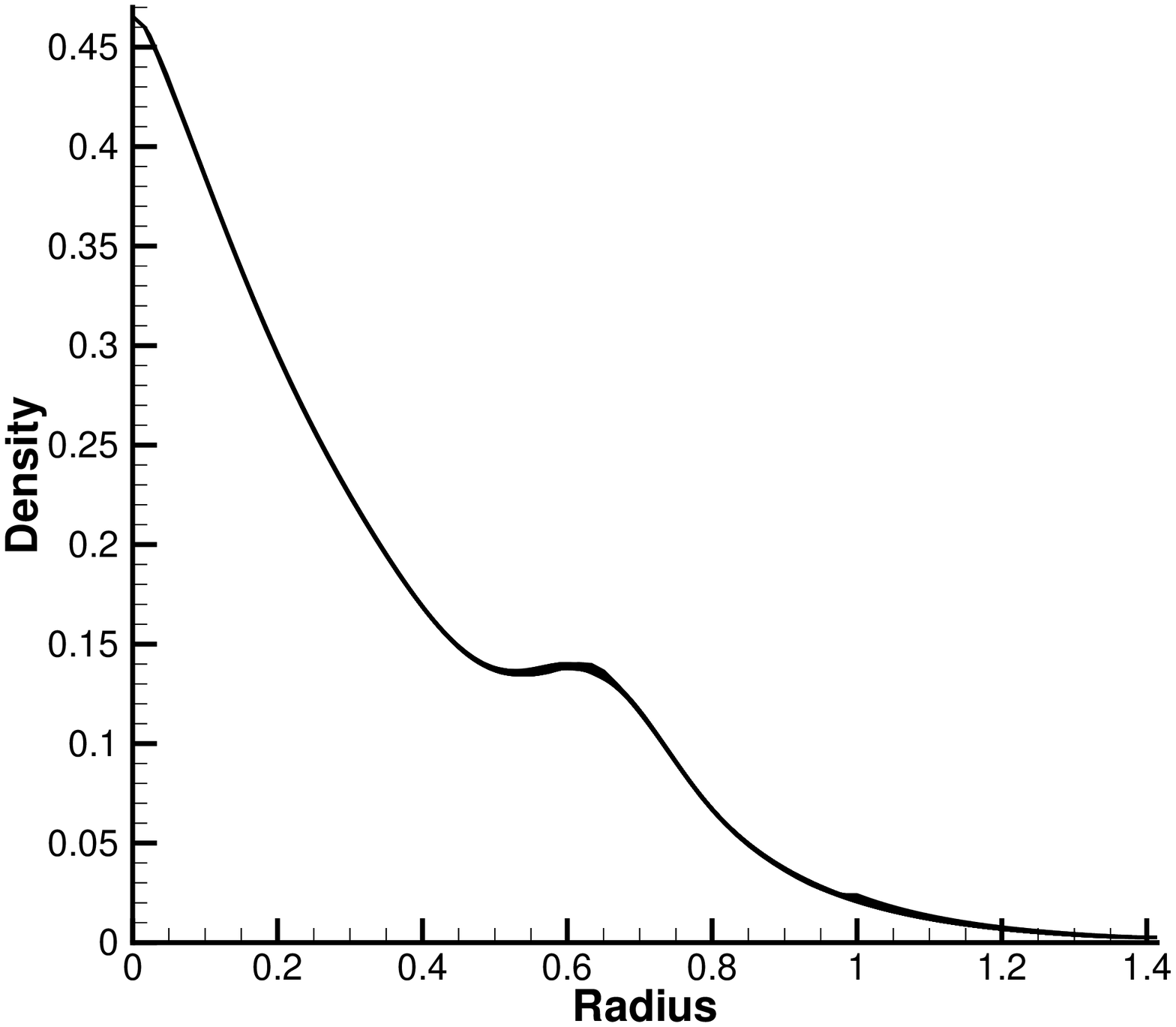}
	}
	\subfigure[t=0.24]{
		\includegraphics[width=3.5cm]{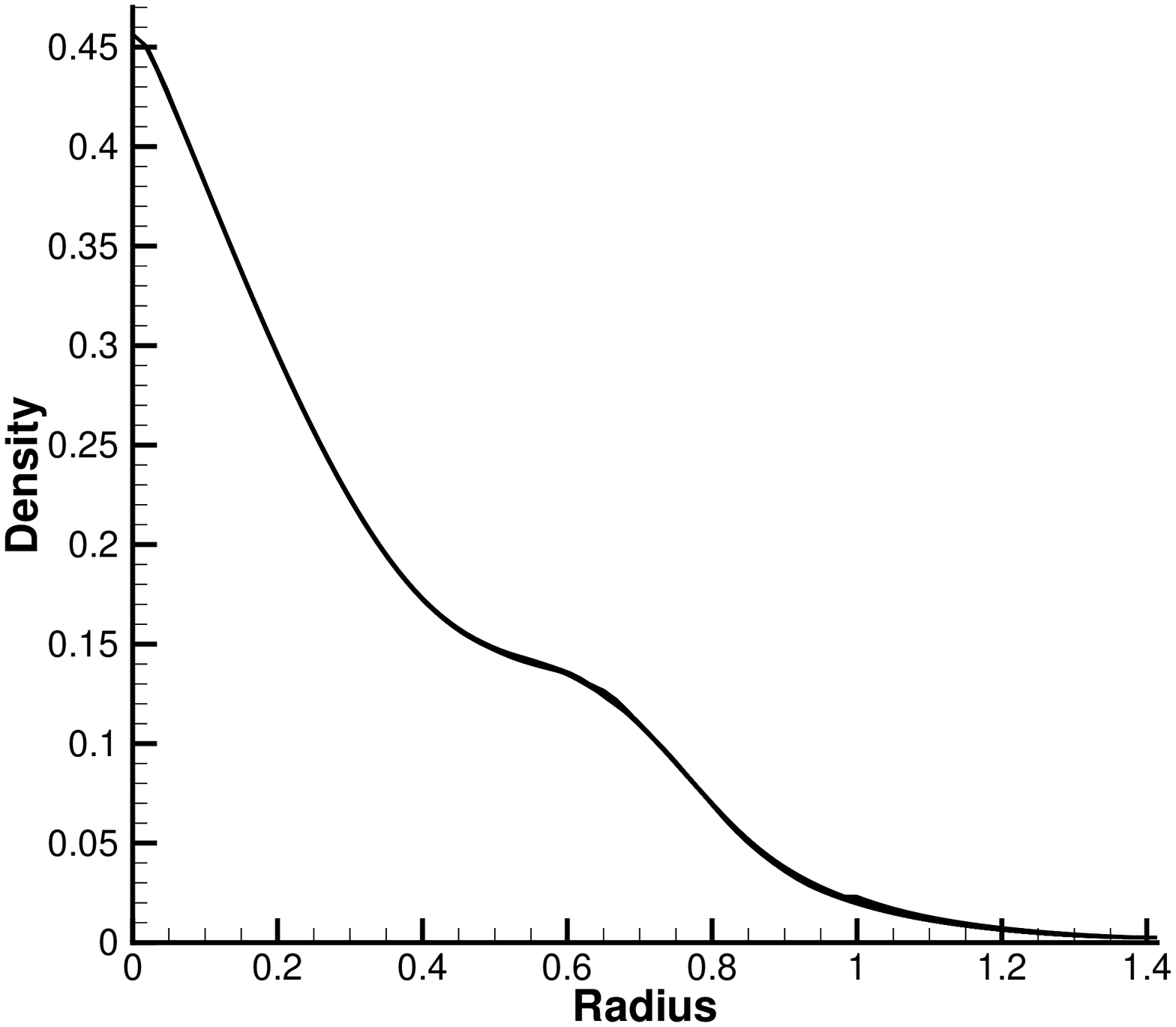}
	}
	\subfigure[t=0]{
		\includegraphics[width=3.5cm]{rtinitialline.eps}
	}
	\subfigure[t=0.08]{
		\includegraphics[width=3.5cm]{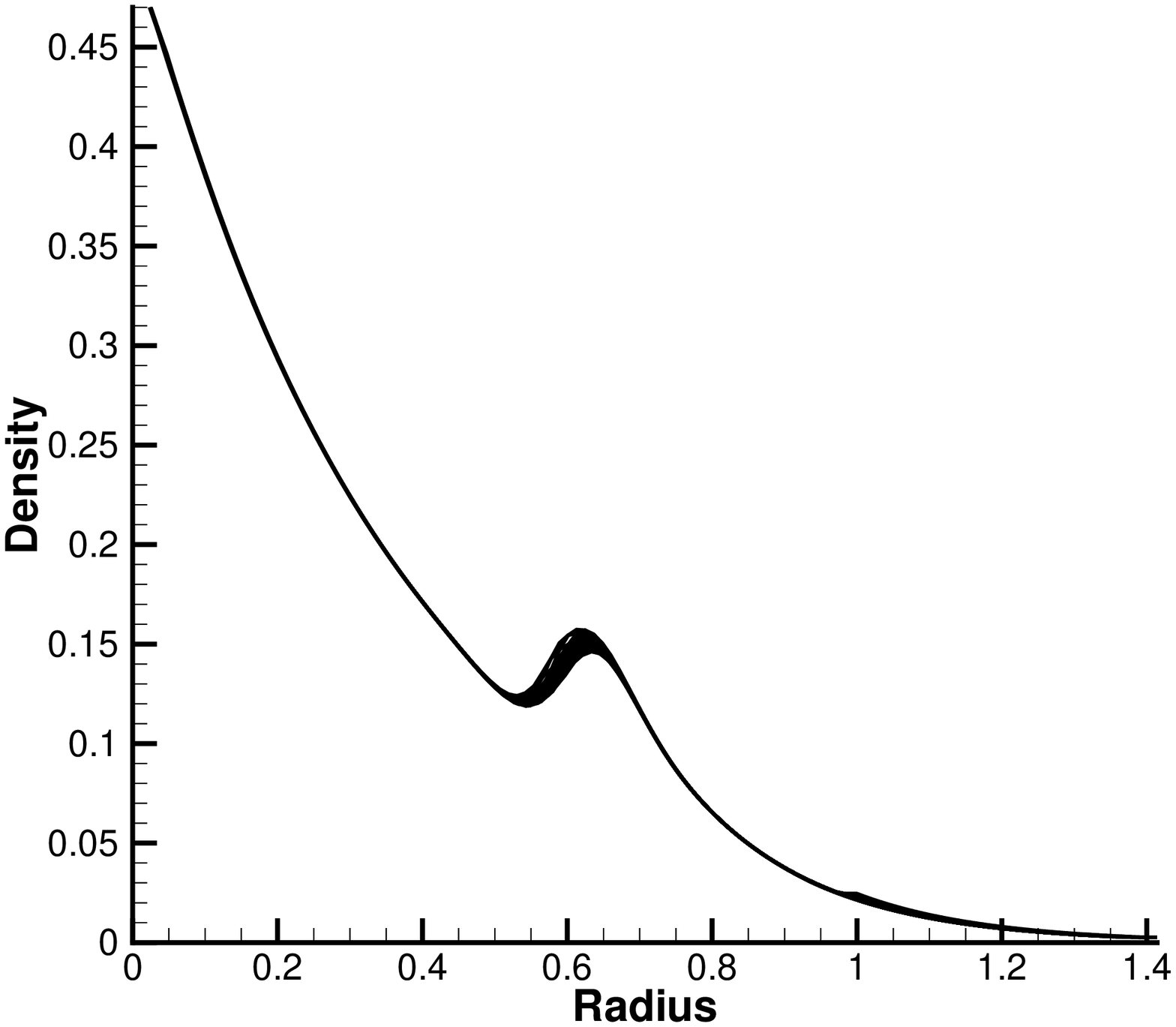}
	}
	\subfigure[t=0.16]{
		\includegraphics[width=3.5cm]{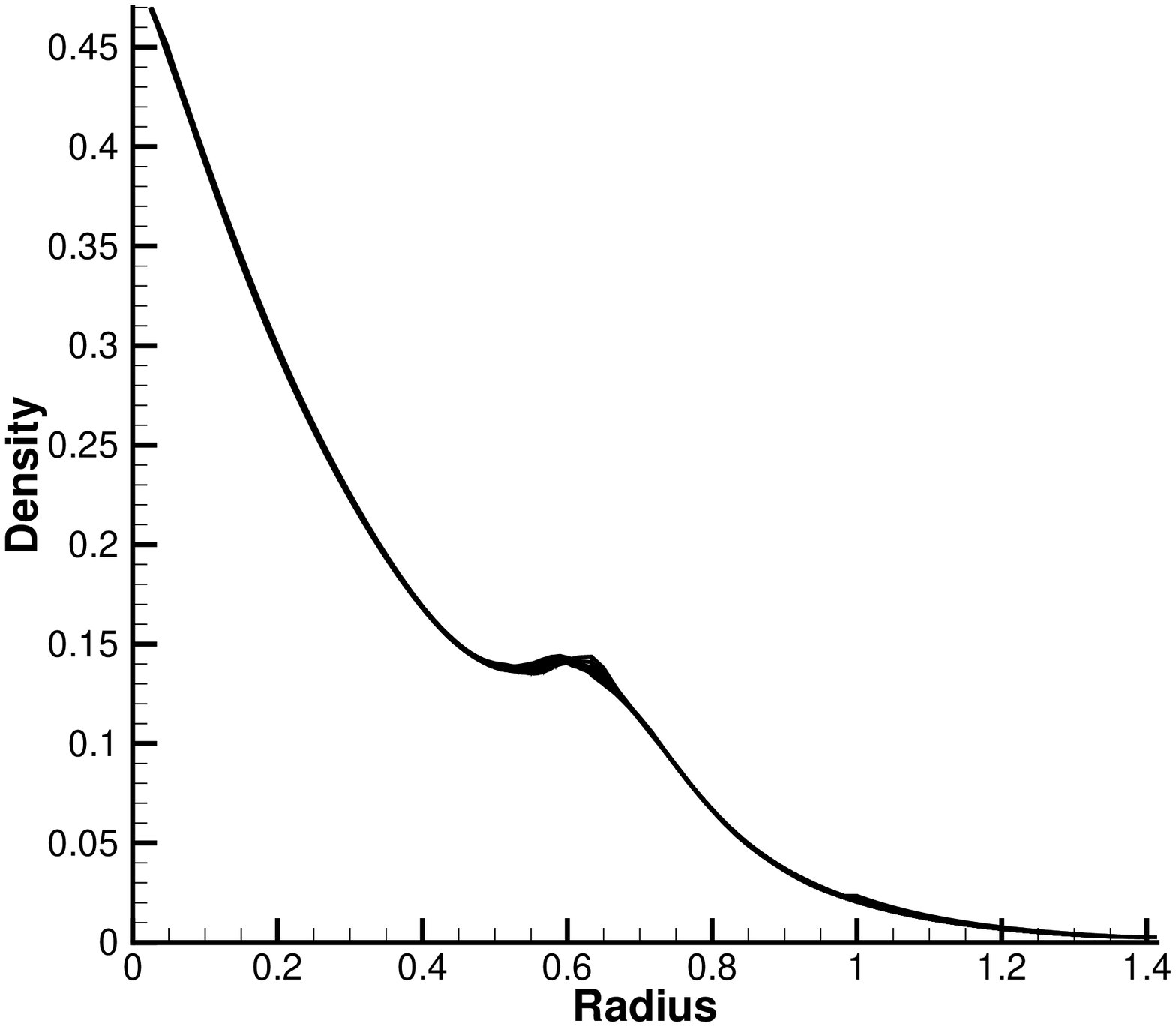}
	}
	\subfigure[t=0.24]{
		\includegraphics[width=3.5cm]{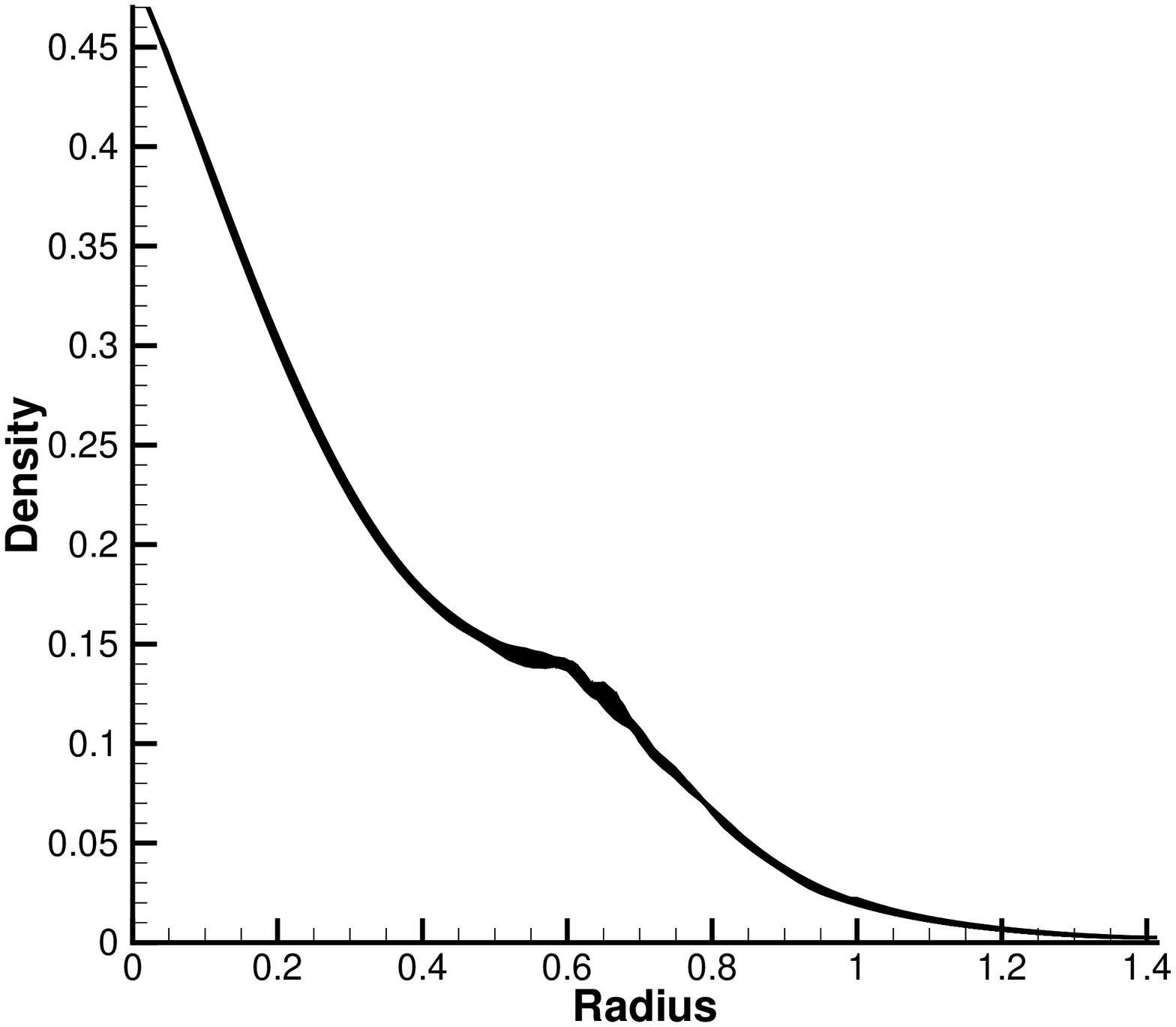}
	}
	\caption{Density distribution along the radial direction with reference Knudsen number 0.0001, 0.01, 1}
	\label{pic:rt line}
\end{figure}

The computational mesh is a $60\times 60$ uniform rectangular one, and the velocity space is divided into $20\times 20$ points.  Different reference Knudsen numbers $\rm Kn=0.0001, 0.01$, and $1$,
are used in the simulation. Here Eq.(\ref{eqn: vss viscosity}) and Eq.(\ref{eqn: hs model}) are employed to evaluated the relation between the dynamic viscosity and the reference Knudsen number,
and the same coefficients for hard-sphere model are used as the shock tube problem. The density contours at different output times are presented in Fig. \ref{pic:rt contour}.
It is clearly demonstrated that the evolution process of Rayleigh-Taylor instability has different features at different rarefaction conditions.
In the continuum regime, as seen in the first row of Fig. \ref{pic:rt contour},
the frequent particle collisions prevent the particle penetration and the strong mixing phenomenon happens in the interface region.
However, as the Knudsen number increases, the mixing processes speed up through the particle penetration, and the phenomenon of interface instability is much weakened.
Fig. \ref{pic:rt line} shows a scattering plot of density for all cells versus the radius from the center in the polar coordinate.
It can be seen that the mixing region in the continuum regime is much narrower than that in the transition and free molecular regimes.
Due to the well-balanced property of UGKS, the hydrostatic solution is well kept in the computation, and the mixing process only occurs near the Rayleigh-Taylor unstable interface.

\subsection{Lid-driven cavity under gravity}

The lid-driven cavity problem is a complex system including boundary effect, shearing structure, heat transfer, non-equilibrium thermodynamics, etc.
In this case, we calculate a multi-scale cavity problem under gravity.
This test case is an ideal one to validate multi-scale methods.

\begin{figure}[htb!]
	\centering
	\subfigure[Density]{
		\includegraphics[width=6cm]{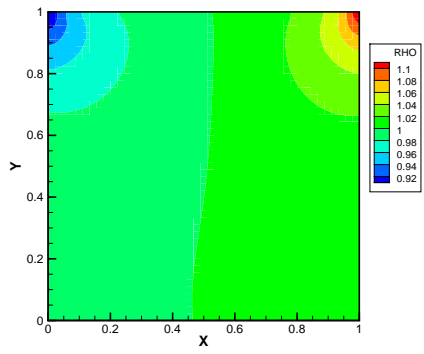}
	}
	\subfigure[Temperature and heat flux]{
		\includegraphics[width=6cm]{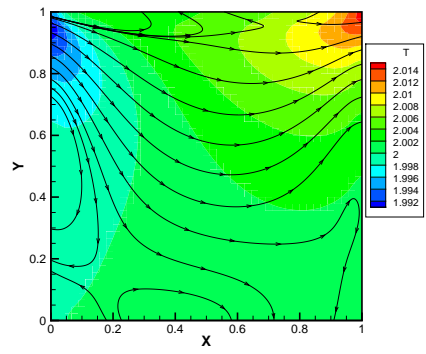}
	}
	\subfigure[U-velocity]{
		\includegraphics[width=6cm]{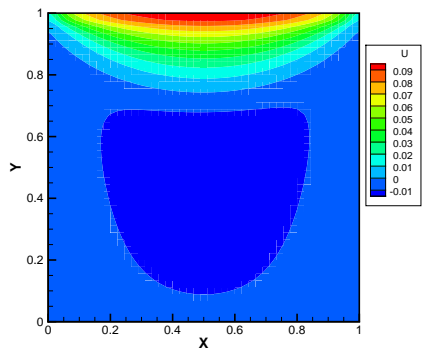}
	}
	\subfigure[V-velocity]{
		\includegraphics[width=6cm]{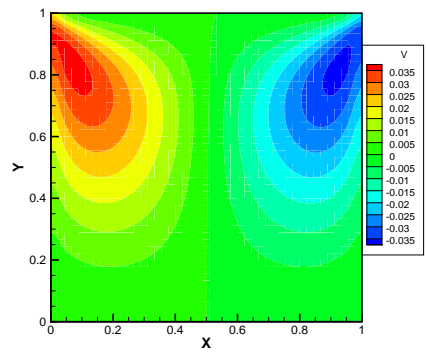}
	}
	\caption{Cavity without gravitational field}
	\label{pic:cavity g=0}
\end{figure}

\begin{figure}[htb!]
	\centering
	\subfigure[Density]{
		\includegraphics[width=6cm]{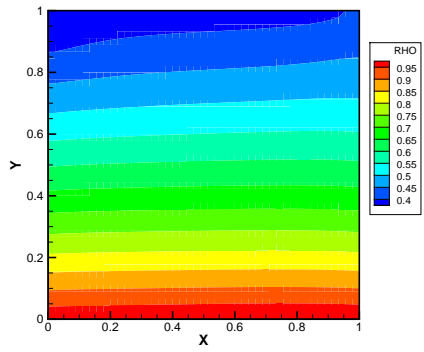}
	}
	\subfigure[Temperature and heat flux]{
		\includegraphics[width=6cm]{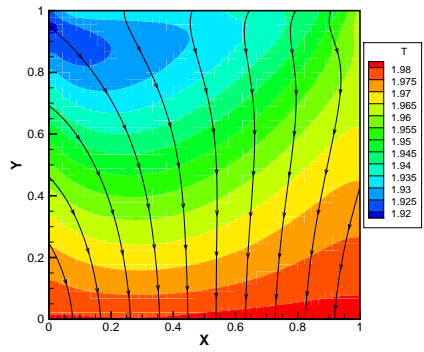}
	}
	\subfigure[U-velocity]{
		\includegraphics[width=6cm]{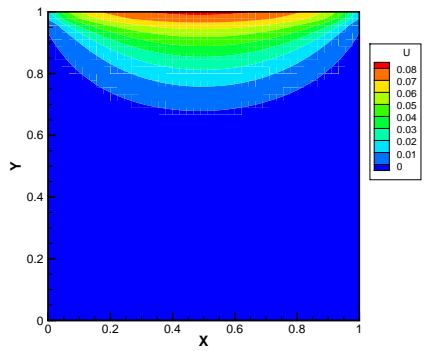}
	}
	\subfigure[V-velocity]{
		\includegraphics[width=6cm]{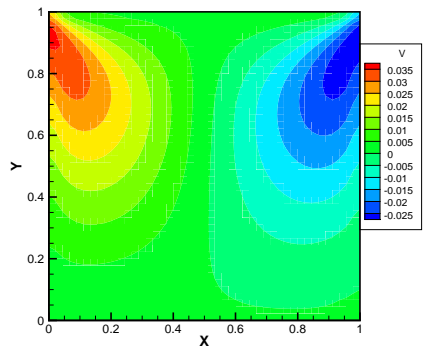}
	}
	\caption{Cavity at $\phi_y= -1.0$}
	\label{pic:cavity g=1}
\end{figure}

\begin{figure}[htb!]
	\centering
	\subfigure[Density]{
		\includegraphics[width=6cm]{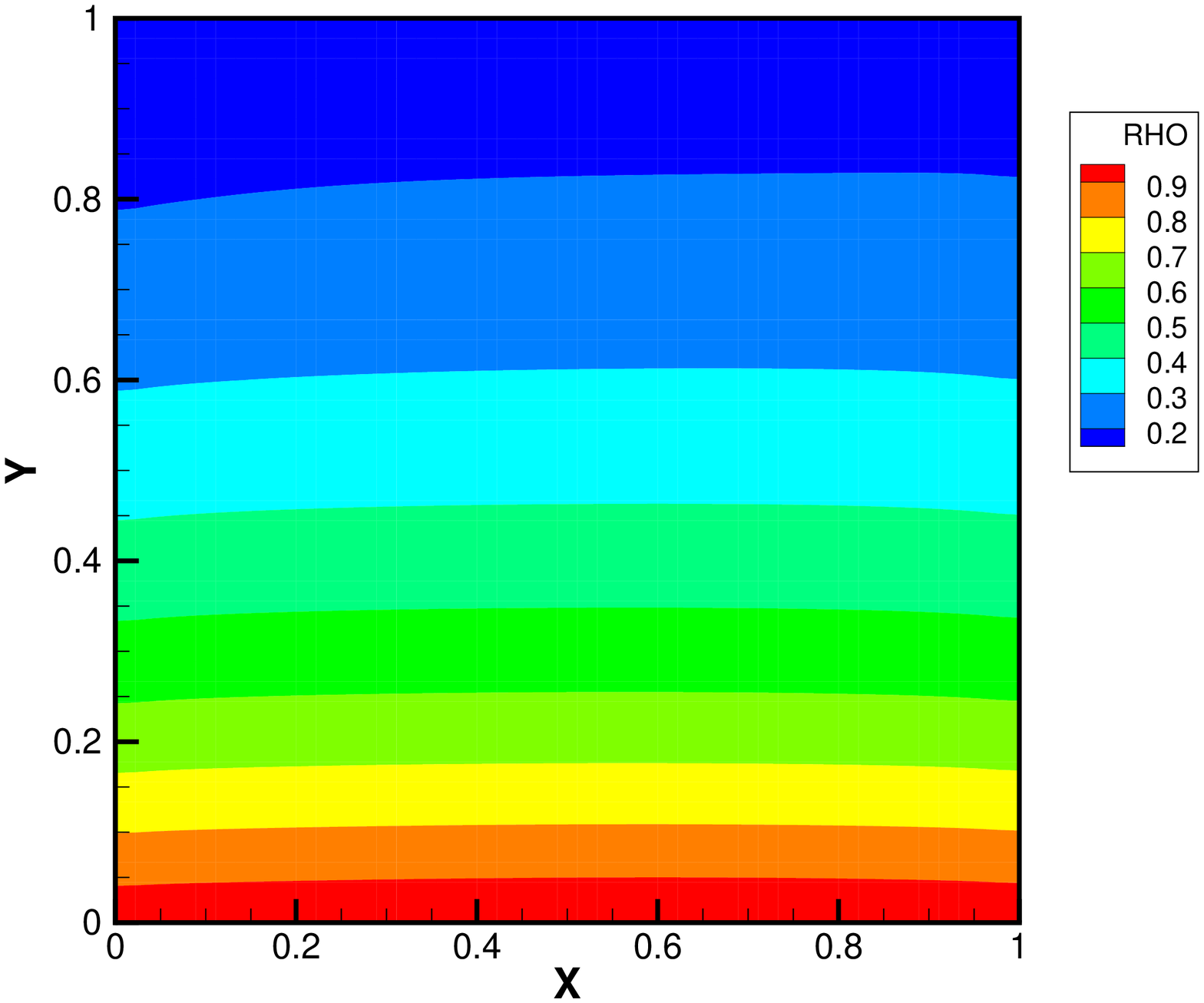}
	}
	\subfigure[Temperature and heat flux]{
		\includegraphics[width=6cm]{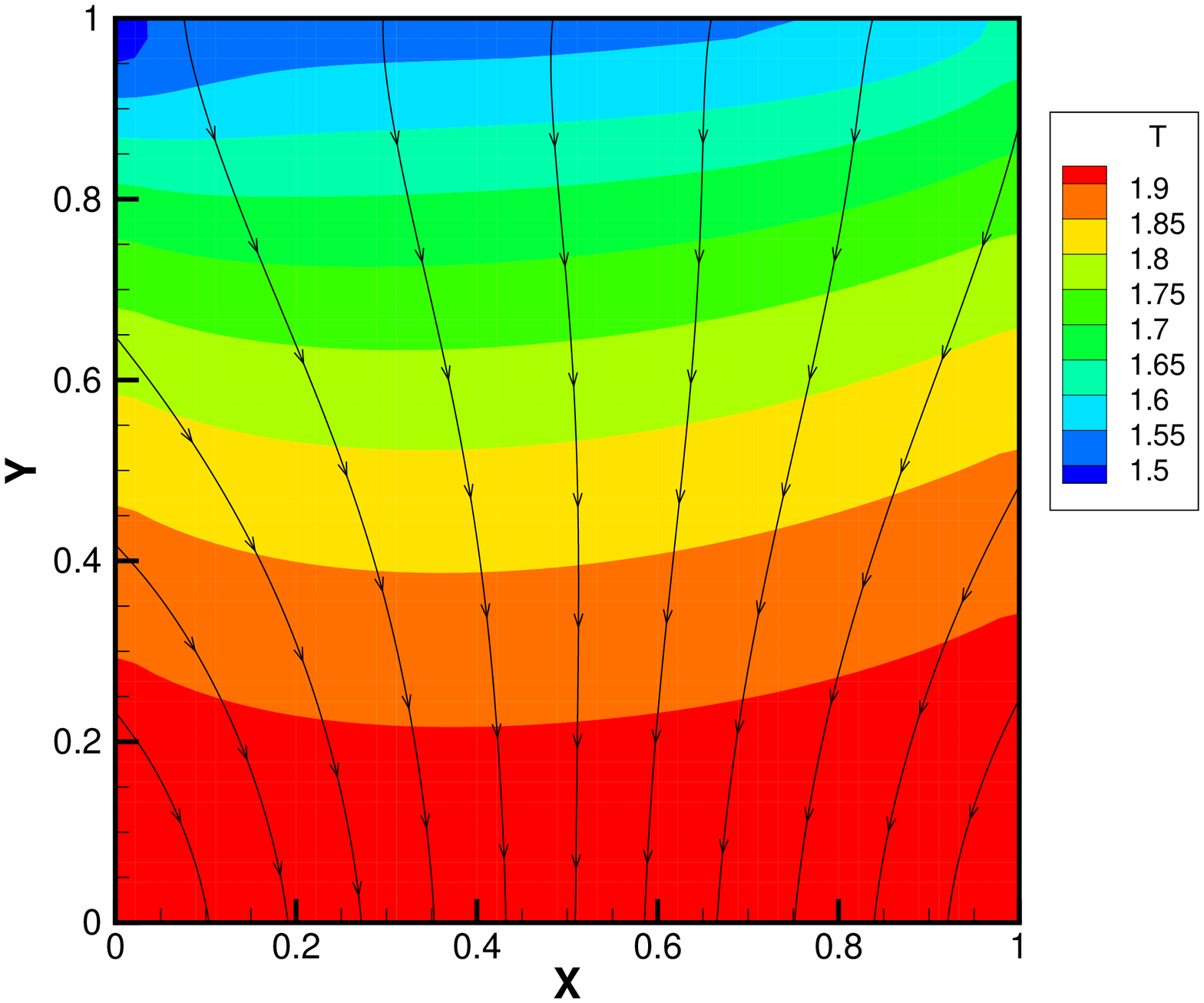}
	}
	\subfigure[U-velocity]{
		\includegraphics[width=6cm]{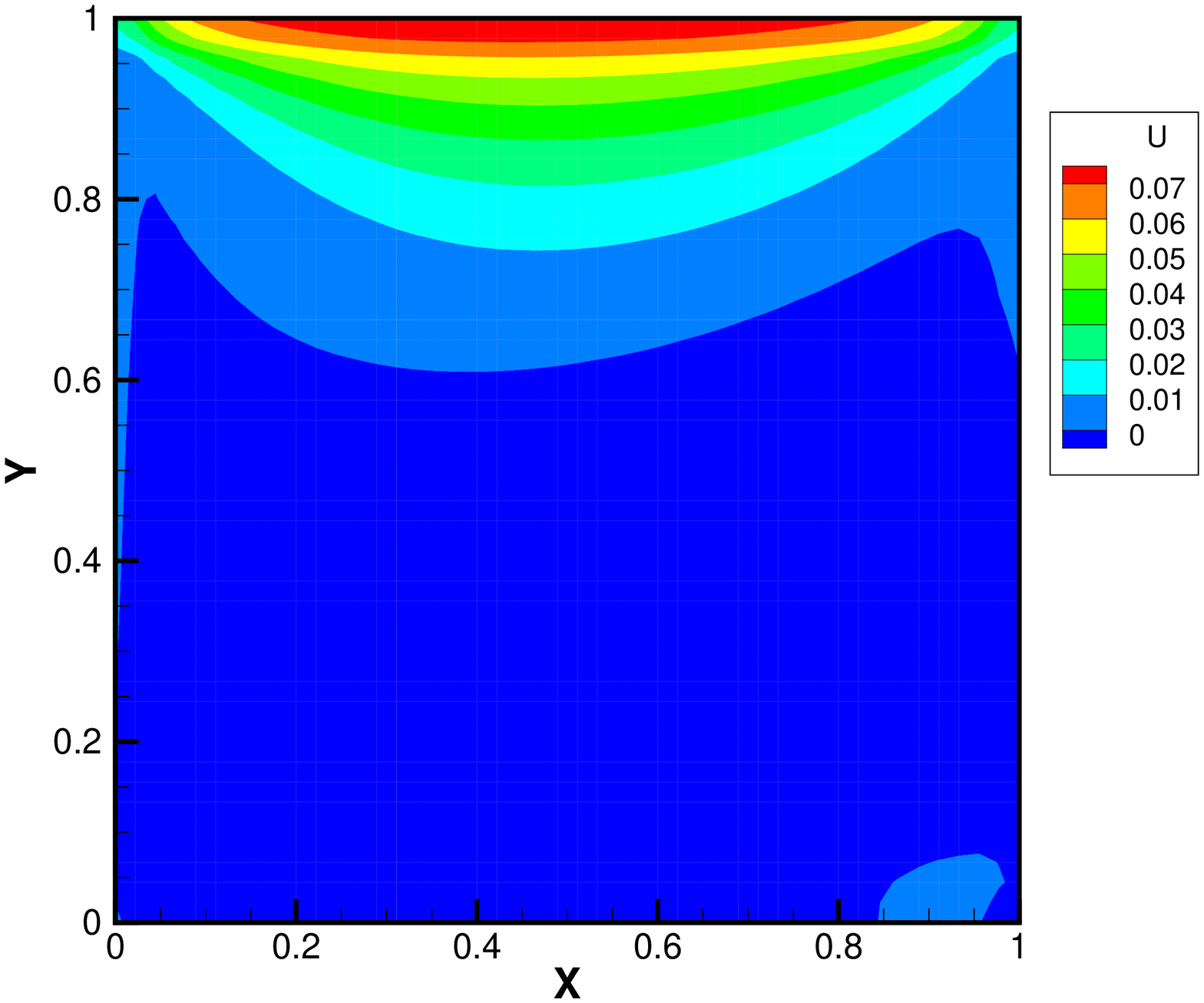}
	}
	\subfigure[V-velocity]{
		\includegraphics[width=6cm]{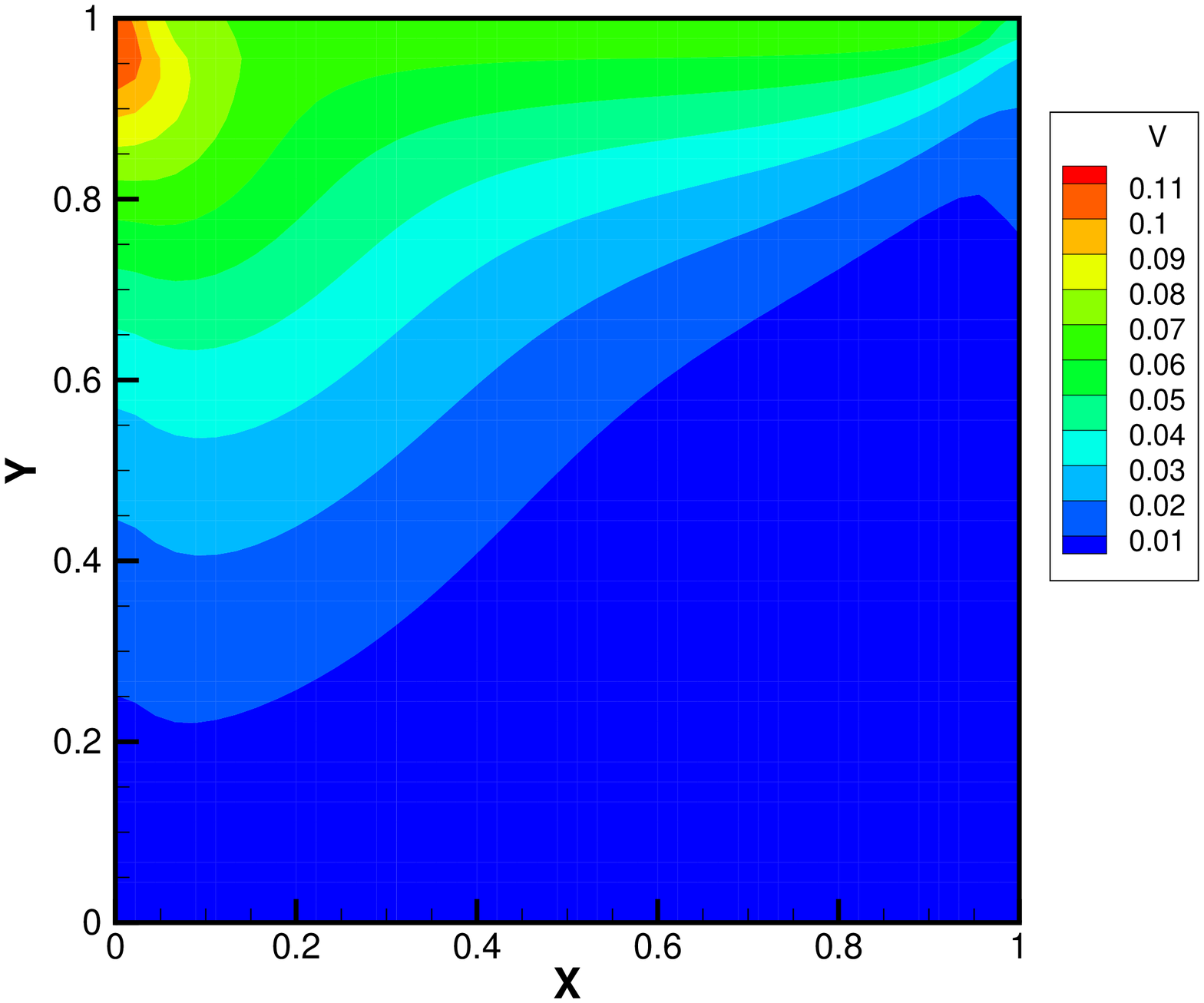}
	}
	\caption{Cavity at $\phi_y= -2.0$}
	\label{pic:cavity g=2}
\end{figure}

The square cavity has four walls with $L=1$. The upper wall moves in tangential direction with a velocity $U_w=0.15$.
The gravity is set to be $\phi_y=-1.0, -2.0$ respectively in the negative $y-$direction. The magnitude of gravity $\phi_y$ is denoted by $g$. The initial density and pressure are set up with
\begin{equation*}
\rho(x,y,t=0)=\exp(\phi_y y),p(x,y,t=0)=\exp(\phi_y y)
\end{equation*}
and wall temperature is $T_w=2$. Maxwell's accommodation boundary condition is used in the simulation. The Prandtl number of the gas is $\rm Pr=0.66667$. The cavity flow under the same initial and boundary condition with the absence of gravity $\phi_y=0.0$, is also simulated using the original unified gas-kinetic scheme for a thorough demonstration.
The reference Knudsen number is defined by reference state at bottom of the cavity $\rho_{\rm ref}=1.0$ and $p_{\rm ref}=1.0$.
The local mean free path $l$ can be evaluated by
\begin{equation}
l_{local}=\frac{\mu}{\mu_{ref}}\sqrt{\frac{p_{ref}}{p}\frac{\rho_{ref}}{\rho}}l_{ref}.
\end{equation}
The reference Knudsen number $\rm Kn=0.075$ is used in the current calculation.
The computational domain is divided into $45\times45$ uniform cells, and a $28\times28$ Gaussian velocity space is employed.
Fig. \ref{pic:cavity g=1}, \ref{pic:cavity g=2} present the simulation results of the two cases under gravitational field.
Fig. \ref{pic:cavity g=0} gives the simulation results of the cavity flow with the absence of gravity.

In the case under gravitational field, the movement of upper wall and gravitational force are two driving sources for the flow motion in the cavity system.
The initial hydrostatic density distribution is perturbed from the upper wall movement.
Due to the gravitational force, the density changes significantly along the vertical direction, so is the local  Knudsen number.
In other words, for this single case, the upper and lower part gas in the cavity may stay in different flow regimes.
As demonstrated, even with the viscous heating at the upper wall, the temperature of the gas around the upper surface of the cavity decreases
due to the energy exchange among gravitational potential energy, kinetic energy, and internal energy.
In comparison with the simulation results without gravitational field, i.e., Fig. \ref{pic:cavity g=0},
the heat under gravity transports in the gravitational field direction from the cold (upper) to the hot (lower) regions.
The cooling of the upper region may have similar mechanism as the dynamic cooling of in the upper atmosphere of the earth.
This observation is against Fourier's law, which is also different from the non-equilibrium heat flux due to the rarefaction effect only \cite{john2011effects}, such as the case
in Fig. \ref{pic:cavity g=0}.
This is the first time that the effect of the gravity on the heat flux has been observed quantitatively. This is a fully non-equilibrium phenomenon.
The results show that the heat flux is  from the cold to the hot region due to the gravitational effect,  which may be used to explain the
gravity-thermal instability in astrophysics.
Different from the equilibrium thermodynamics, the shift and distortion of the gas distribution function due to the external forcing term
provide the dominant mechanism for the non-equilibrium heat flux, especially in the transition flow regime with modest Knudsen number.

\begin{figure}[htb!]
	\centering
	\subfigure[U-velocity along the vertical center line]{
		\includegraphics[width=6cm]{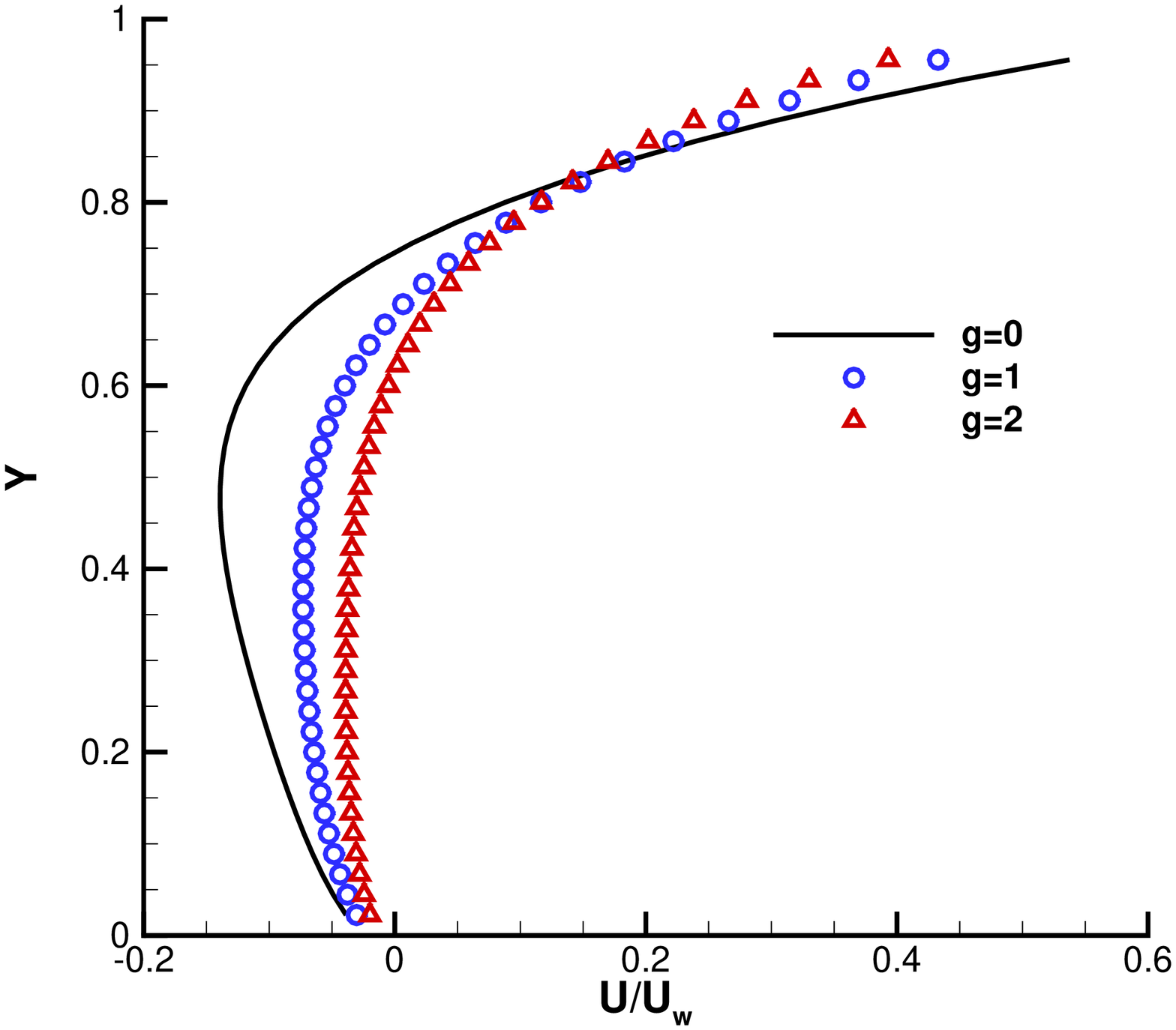}
	}
	\subfigure[V-velocity along the horizontal center line]{
		\includegraphics[width=6cm]{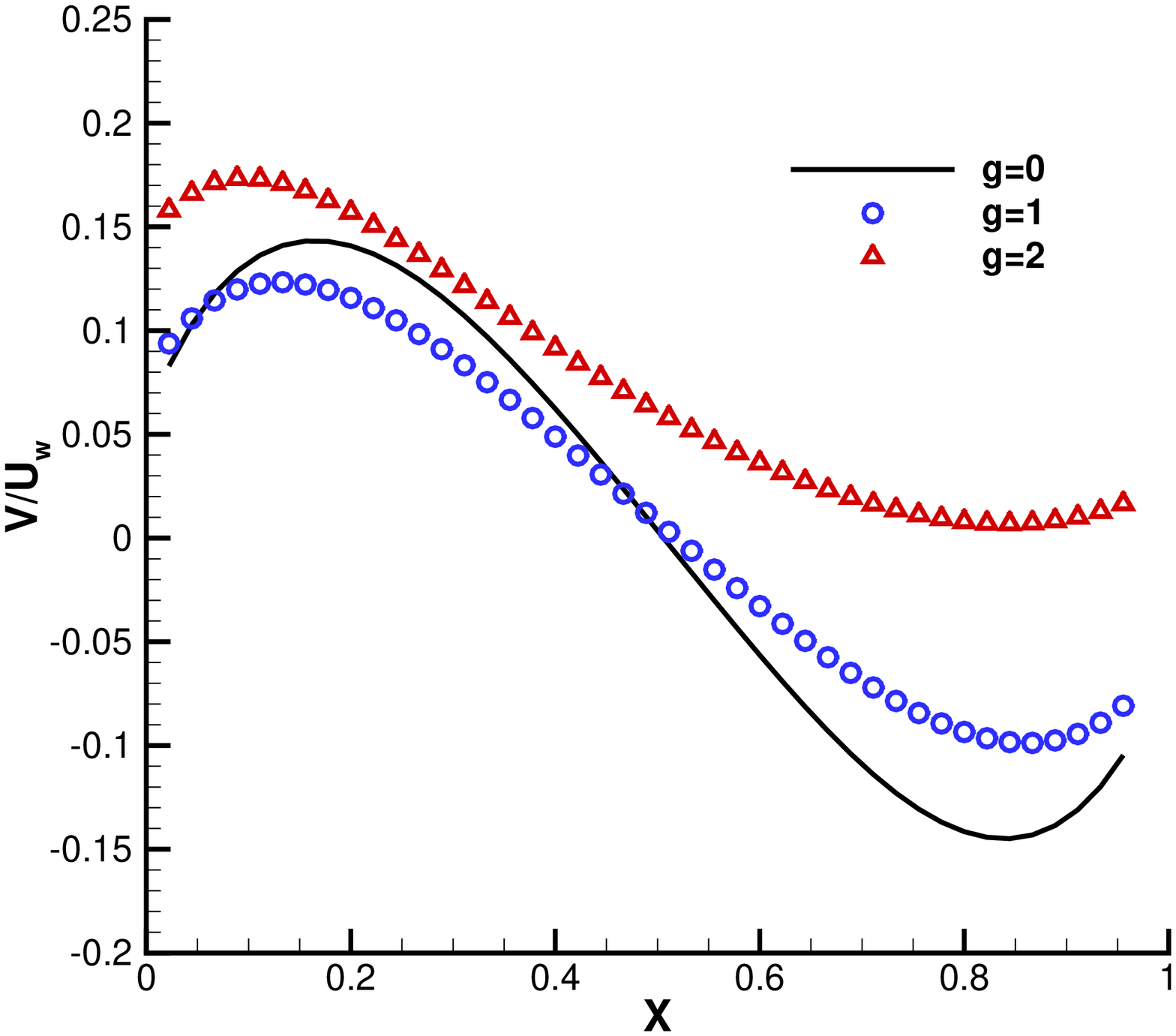}
	}
	\caption{U,V velocity at the horizontal and vertical center line}
	\label{pic:cavity velocity curve}
\end{figure}

\begin{figure}[htb!]
	\centering
	\includegraphics[width=6cm]{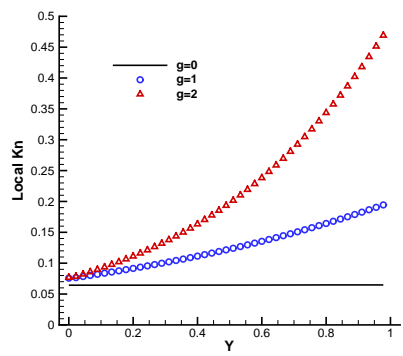}
	\caption{Local Knudsen number at the vertical center line}
	\label{pic:cavity local kn}
\end{figure}

The U-velocity and V-velocity distributions at the central vertical and horizontal line of the cavity as well as the local Knudsen number are presented
in Fig. \ref{pic:cavity velocity curve} and Fig. \ref{pic:cavity local kn}.
As presented in Fig. \ref{pic:cavity velocity curve}, with the increment of gravitational force,
the fluid motion is mostly controlled by gravity. In the case with $g=2$, the local Knudsen number changes significantly from $0.5$ at top of the cavity to $0.075$ at the  bottom of the cavity.
The multi-scale flow physics will appear inside the cavity. This case clearly demonstrates the capacity of the unified scheme to study highly complicated non-equilibrium flow phenomena.

\section{Conclusion}

Gas dynamics under gravitational field has a multiple scale nature with large variation of gas density in different regions.
Based on the direct modeling, a well-balanced unified gas-kinetic scheme under gravitational field has been constructed in this paper for the flow simulation in all regimes.
The well-balanced property of the scheme for converging and maintaining a hydrostatic equilibrium solution is validated through numerical tests.
At the same time, due to the multiple scale modeling the UGKS can capture the non-equilibrium flow phenomena associated with the gravitational field.
The UGKS is a reliable algorithm for flow simulations from continuum to rarefied one.
In the transition regime, for the first time the non-equilibrium flow phenomenon, such as the correlation between the heat flux and gravitational field,
has been observed  in the cavity case. This scheme may help the modeling for the large-scale atmospheric flow around earth surface, and the quantitative study of gravity-thermal instability.
The well-balanced UGKS provides an indispensable tool for the study of the multiple scale non-equilibrium gas dynamics under gravitational field.

\section*{Acknowledgement} 
The current research  is supported by Hong Kong research grant council (16207715, 16211014, 620813), and  National Science Foundation of China (91330203,91530319).

\clearpage

\bibliographystyle{unsrt}
\bibliography{gksforce}


\end{document}